\documentclass[manuscript,screen,nonacm,oneside]{acmart}

\usepackage{fontspec}
\usepackage{subcaption}
\newfontface{\bn}{kalpurush.ttf}
\usepackage{multirow} 
\usepackage{enumitem}
\AtBeginDocument{%
  }

\copyrightyear{2018}
\acmYear{2018}
\acmDOI{XXXXXXX.XXXXXXX}

\begin{document}

\title{A Usable and Secure Bengali CAPTCHA}

\author{Md Neyamul Islam Shibbir}
\authornote{Both authors contributed equally to this research.}
\email{mshibbir@miners.utep.edu}
\orcid{0009-0005-6237-2859}
\affiliation{%
  \institution{The University of Texas at El Paso}
  \city{El Paso}
  \state{Texas}
  \country{USA}
}

\author{Md Hasibur Rahman}
\authornotemark[1]
\orcid{0009-0001-0439-2449}
\email{hamim4389@gmail.com}
\affiliation{%
  \institution{Shahjalal University of Science and Technology}
  \city{Sylhet}
  \country{Bangladesh}}

\author{Farida Chowdhury}
\orcid{0000-0001-9902-6291}
\email{farida.chowdhury@bracu.ac.bd}
\affiliation{%
  \institution{BRAC University}
  \city{Dhaka}
  \country{Bangladesh}
}

\author{Md Sadek Ferdous}
\orcid{0000-0002-8361-4870}
\email{sadek.ferdous@bracu.ac.bd}
\affiliation{%
  \institution{BRAC University}
  \city{Dhaka}
  \country{Bangladesh}
}

\renewcommand{\shortauthors}{MNI Shibbir et al.}

\begin{abstract}
  Text-based CAPTCHAs (Completely Automated Public Turing test to tell Computers and Humans Apart) have traditionally been a simple, affordable, lightweight, yet very effective security mechanism to distinguish human users from automated bots on the web, serving as a preventive measure against many cyberattacks. However, the dependence on the English script creates usability issues for non-native speakers, limiting accessibility for regional communities where English is not widely understood. In this work, we have proposed and implemented a text CAPTCHA mechanism with 6 variants on the Bengali language, designed specifically for native Bengali-speaking users, which is the first of its kind to the best of our knowledge. Our proposed Bengali CAPTCHA exhibits robust security against automated OCR-based attacks, limited to only 0–20\% average character recognition rate across 6,000 challenges (1,000 per variant approx.). Furthermore, our design demonstrates high human usability, evaluated with 110 participants, achieving success rates of 56.25\% to 90.29\% and average response times of 6.69 to 9.9 seconds across all six variants, thereby standing out among text-based CAPTCHA benchmarks. 
\end{abstract}




\maketitle
\fancyfoot[RE,LO]{}

\section{Introduction}
The Human Interaction Proofs (HIP) is often regarded as the most widespread security defense mechanism against automated bots, providing reliable assurance that the a client accessing remote services is a human being rather than bot mimicking human behavior \cite{chellapilla2005computers, belk2015human}. Completely Automated Public Turing test to tell Computers and Humans Apart (CAPTCHA) \cite{CAPTCHA} is one type of HIPs that tests agents by providing problems which are easily solvable by humans, but will be very difficult to solve for bots. CAPTCHA tests are considered valid and successful if the tests can be solved by humans with a success rate over 90\% and the success rate of automated software or bots less than 0.1\% \cite{brodic2019captcha}. 

CAPTCHA plays a decisive role in websites nowadays. Nevertheless, it also comes with various security concerns. 
Most existing CAPTCHA systems are primarily available in English, which limits accessibility for users from regional communities with limited proficiency in English. To address this gap, we propose a secure and usable Bengali CAPTCHA designed for regional websites targeting Bengali speakers. The proposed Bengali CAPTCHA is intended for Bengali-first digital services where users interact primarily in Bengali script. In such contexts, using a Bengali CAPTCHA can make the verification process more familiar and consistent with the rest of the interface. This is especially relevant for local government portals, educational platforms, banking services, regional e-commerce websites, and other Bengali-oriented online services, where presenting the CAPTCHA in the same language as the surrounding content may improve user convenience and accessibility. The research challenge is not only translating an English CAPTCHA into Bengali, but adapting the CAPTCHA design space to the linguistic and technical properties of Bengali script. Bengali contains visually similar characters, vowel diacritics, conjunct forms, and input-method constraints that directly affect both human readability and automated recognition. Therefore, designing a Bengali-native CAPTCHA requires script-aware character selection, exclusion of confusing or difficult-to-render characters, and evaluation of how Bengali-specific visual forms interact with common CAPTCHA distortions. Although this study does not directly measure the failure rate of English CAPTCHAs among Bengali-speaking users, the large Bengali-speaking population and the growing number of Bengali-first digital services motivate the need for localized verification mechanisms. The proposed system therefore provides an initial empirical baseline for Bengali-native CAPTCHA design rather than claiming to replace English CAPTCHAs in all contexts. According to Statista \cite{BengaliMost}, Bengali is the seventh most spoken language in the world with 284.3 million speakers. Furthermore, no prior CAPTCHA system has been developed in Bengali, there is a clear need for such mechanisms.  
 
To a certain extent, the use of a regional language introduces a linguistic barrier for automated bots, as breaking such CAPTCHAs requires advanced Bengali OCR capabilities. This approach leverages the current limitations of Bengali OCR systems in accurately recognizing Bengali texts. However, CAPTCHA design inherently involves balancing security and usability, and identifying an optimal trade-off between the two. In this article, we introduce a novel text-based CAPTCHA for Bengali users that achieves both security and usability.

\noindent \textbf{Contributions:} The major contributions of this article are:
\begin{itemize}
  \item We have designed and implemented a secure and usable CAPTCHA scheme for Bengali language leveraging appropriate methods and elements of Bengali linguistics and state-of-the-art of image processing tools and techniques. 
  \item We have analyzed and tested the security of the implemented CAPTCHA scheme using a recently developed evaluation framework for text-based CAPTCHAs \cite{shibbir2024evaluating}.
  \item We have conducted an IRB-approved usability study with 110 native Bengali speakers and introduced a normative-comparison methodology that ranks CAPTCHA variants against an ideal usability profile using RMSE and correlation.

\end{itemize}

\section{Background}
\label{sec:background}
CAPTCHA (Completely Automated Public Turing test to tell Computers and Humans Apart) is a widely adopted mechanism designed to distinguish human users from automated bots accessing online services. Introduced in 2000 \cite{brodic2019captcha}, CAPTCHA systems present tasks that are easily solvable by humans but challenging for computers, typically relying on human cognitive skills such as visual perception and pattern recognition \cite{brodic2019captcha, chellapilla2005computers}. Originally conceptualized as reverse Turing tests \cite{brodic2019captcha, ahn2003captcha}, CAPTCHAs automate the process of verification where the computer judges whether the user is human based on their ability to complete specific tasks.

Various forms of CAPTCHA have been developed, including text-based, image-based, audio, and game-based types \cite{carlos2010Pitfalls, manar2014, oleg2015, yan2008usability}. Text-based CAPTCHAs remain one of the most prevalent due to their intuitive design and ease of automation \cite{yan2008usability, chow2019captcha}. However, with rapid advancements in artificial intelligence and computer vision, such CAPTCHAs face increasing threats from sophisticated algorithms capable of bypassing traditional challenges. This has resulted in an ongoing arms race between CAPTCHA developers and attackers, with continued research focusing on enhancing security without compromising user experience \cite{chellapilla2005designing, chellapilla2005computers}.

According to \cite{brodic2019captcha} there are three important factors to consider while designing a CAPTCHA mechanism:
\begin{itemize}
    \item \textbf{Security}: The security element defines the process to protect the CAPTCHA from being attacked, 
    \item \textbf{Usability}: The usability element defines the solvability of CAPTCHAs by the human users,
    \item \textbf{Practicality}: Finally, the practicality element defines the feasibility of the mechanism to be implemented on a website. 
\end{itemize}
To enhance CAPTCHA security, researchers can incorporate numerous security characteristics, such as deformation, color variation, rotation, blurring, warping, multi-level structure, overlapping characters, and noisy backgrounds, to make them more challenging to crack \cite{shibbir2024evaluating}. For example, text CAPTCHAs typically use distorted words to increase security. However, using certain letters in the English language is discouraged. These letters include 6/G, b/5, S/s, O/0, Z/z, nn/m, w/vv, d/cl and others since they can be difficult for real humans to recognize and separate after being distorted \cite{ahmmed2014, yan2008confusing}. Even so, typical text CAPTCHAs can easily be broken using OCR (Optical Character Recognition) techniques \cite{eikvil1993optical}. Examples of some text CAPTCHAs are Gimpy \cite{mori2013captcha}, EZ-Gimpy \cite{kaur2014captcha}, Baffle Text \cite{researchgate} and Pessimal Print \cite{powell2010image}.

\section{Related Work}
\label{chap:relatedWork}
To the best of our knowledge, there are no Bengali CAPTCHAs at the time of this research. However, linguistic CAPTCHAs are not uncommon in the research field. Kumar et al. \cite{kumar2022design} broke the existing 20 Hindi Text CAPTCHAs and came up with a more secure and usable Hindi CAPTCHA. In addition, the authors recommended some techniques and features that must be kept in mind while developing a text CAPTCHA in any language. Furthermore, the authors in \cite{khan2013cyber} used specific Arabic font types to develop an Arabic Text CAPTCHA, exploiting the limitations of Arabic OCR techniques. They tested their CAPTCHA against Arabic OCRs but failed to recognize the CAPTCHA text. Lastly, they conducted a user study that showed that their developed CAPTCHA is also user-friendly.

Banday et el. \cite{banday2013design} reviewed the existing CAPTCHA schemes in Indian regional languages and then proposed a scheme that can generate Hindi, Punjabi, Urdu and English languages CAPTCHAs. Nevertheless, no usability study was conducted on the work. In \cite{dahar2020enhancing}, the authors proposed and developed an Urdu text CAPTCHA. Their proposed system challenges users with single, double, meaningless, and meaningful words having text lengths from 4 to 8. They reported that increasing security features inversely reduces the response time. A CAPTCHA scheme in Malay language is discussed in \cite{yamaba2021proposal}. The authors suggested employing the Jawi script for CAPTCHA and more especially, the digraphia feature, which combines two characters to represent a single linguistic sound.

Trong et al. \cite{trong2023new} used a combination of deep learning and cognition on text-based CAPTCHAs. They claim that this process remarkably increases the security of the proposed CAPTCHA scheme. However, they did not provide any usability data or studies to point out how usable the CAPTCHA scheme is to the users. Work on designing a text CAPTCHA that is both secure and usable is still limited. Kaur et al. \cite{KAUR2015122} provided an algorithm to secure text CAPTCHAs, but the usability of the CAPTCHA is not discussed there either. On the other hand, there are some works focusing only on the usability aspects of CAPTCHAs. A three-dimensional framework was developed consisting of Distortion, Content and Presentation of CAPTCHAs to measure the Usability \cite{yan2008usability}. However, the real user input was not considered in that usability study. User inputs were mostly taken by a set of questionnaires in the design and development of another innovative CAPTCHA \cite{chhabra2019design}.  Brodic et al. \cite{brodic2019exploring} used advanced statistical analysis named association rule mining to perform usability analysis that points out relevant information for creating new CAPTCHAs.

\section{Designing a Secure and Usable Bengali CAPTCHA}
\label{section4: design of a secure and usable bengali cAPTCHA}
The goal of designing a CAPTCHA is to present challenges that are easily solvable by humans but difficult for automated bots. Among various CAPTCHA types, text-based CAPTCHAs are the simplest and the most widely used on the Internet \cite{7207116, chow2019captcha}, due to their accessibility, simplicity, effectiveness, and user familiarity \cite{singh2014survey}. Designing a text-based CAPTCHA requires careful consideration of factors such as character selection, fonts, background patterns, and noise. A critical challenge is maintaining an appropriate balance, as excessive noise reduces human usability, while insufficient complexity increases vulnerability to automated attacks. In the following, we discuss key aspects of designing a secure and usable Bengali CAPTCHA.

\subsection{Character Set}
The set of characters to be chosen is the key issue while developing a text CAPTCHA. For clarity and implementation considerations, we must pick the right characters from the whole Bengali character set (as presented in Table \ref{tab:bengali-script}). We have selected 30 characters, based on some important factors discussed below, from the Bengali language and Bengali numbers, as shown in Table \ref{tab:used-removed-letters}.

\begin{itemize}
    \item The Bengali script consists of vowels, vowel diacritics, consonants, consonant conjuncts, diacritical and other symbols, numbers, and punctuation marks (Table \ref{tab:bengali-script}). There are 47 letters and 10 digits in the Bengali script. The letters are broadly categorized into vowels and consonants. Vowel diacritics are special marks used when a vowel follows a consonant. In such cases, the vowel is not written separately; instead, it is represented as a diacritical mark placed around the consonant—often before or attached to it. Consonant conjuncts (also known as conjunct letters or clusters) are formed when two consonants appear together without a vowel in between. In Bengali, as well as in some other Indic scripts, these combinations are merged into a single, visually distinct character. Conjuncts are very common in Bengali writing and are essential to the script’s structure and pronunciation. Among these characters, those that resemble one another after applying distortion, causing confusion, are excluded. We have shown the confusing and excluded letters in Table \ref{tab:used-removed-letters}.
    \item Many keyboards are used to write Bengali texts, some of which accept English letters as input and output Bengali Unicode characters. Others present Bengali Unicode letters on the keyboard, and the user can click these letters to type any Bengali text. Some popular keyboards are the Avro Keyboard \cite{OmicronLab}, Bijoy Bayanno \cite{Bijoy}, and Ekushey \cite{ekushey}. Most Bengali users prefer the keyboard that lets them type English letters as input and get Bengali letters in the output. However, certain characters are more difficult and lengthy to write in English letters. For example, if we want to  type \bn{ঞ, ঙ, ঋ, ঔ, ঐ}, we need to input these English letters \textit{NG, Ng, rri, OU, OI} respectively, which is considered difficult and lengthy. Also, the letters {(\bn{ঔ, ঋ, ঐ, ঊ, ৎ, ং,  ঃ, ঁ, ঞ, ঙ, ঈ})} are scarcely used by Bengali users in texting, as their frequency is quite low in Bengali online contents \cite{9845427}. So, we have removed these letters to improve usability.
    \item During the generation of the CAPTCHA texts, we discovered that most of the vowel diacritics do not match the input text in the output text-image (Table \ref{tab:vowel-problems}). Therefore, the user will get an invalid or wrong answer if this CAPTCHA is employed. That is why we have omitted these vowel diacritics from our CAPTCHA texts.
    \item We have also eschewed the use of diphthongs or compound letters since their depiction breaks apart during image transformation (Table \ref{tab:conjunct-problems}). Furthermore, we were unable to utilize {(\bn{`(রেফ), ্ (হষন্ত)})} as diacritics since they get mixed with other noises.
\end{itemize}

\begin{table}[h]
\centering
\small
\setlength{\tabcolsep}{4pt}
\begin{tabular}{l p{0.22\linewidth} p{0.6\linewidth}}
\toprule
\textbf{Category} & \textbf{Subtype (Count)} & \textbf{Characters} \\
\midrule \midrule
Letters & Consonants-36 &
\bn{ক, খ, গ, ঘ, ঙ, চ, ছ, জ, ঝ, ঞ, ট, ঠ, ড, ঢ, ণ, ত, থ, দ, ধ, ন, প, ফ, ব, ভ, ম, য, র, ল, শ, ষ, স, হ, ড়, ঢ়, য়, ৎ} \\

& Vowels-11 &
\bn{অ, আ, ই, ঈ, উ, ঊ, ঋ, এ, ঐ, ও, ঔ} \\
\midrule
Diacritics & Vowel signs-10 &
\bn{া, ি, ী, ু, ূ, ৃ, ে, ৈ, ো, ৌ} \\

& Consonant signs-7 &
\bn{ং, ঃ, ঁ, ্, ্য, ্র, `} \\
\midrule
Conjuncts & Examples &
\bn{ন্ধ, ন্ড, ন্দ, ন্ত, ন্ঠ, দ্দ, ক্ল, ত্ন, ন্ত্র, গ্ধ, \ldots} \\
\bottomrule
\end{tabular}
\caption{Overview of Bengali script components \cite{wikipedia_2023}}
\label{tab:bengali-script}
\end{table}

\begin{table}[h]
\centering
\small
\setlength{\tabcolsep}{4pt}
\begin{tabular}{l p{0.22\linewidth} p{0.6\linewidth}}
\toprule
\textbf{Type} & \textbf{Reason / Subtype} & \textbf{Characters} \\
\midrule \midrule
Used & Chosen letters and digits (30) &
\bn{ক, খ, গ, ঘ, চ, ছ, জ, ঝ, প, ফ, ঠ, ভ, ম, ল, শ, স, হ, ষ, অ, আ, ই, উ, এ, ও, ১, ২, ৫, ৬, ৭, ৯} \\
\midrule
\multirow{5}{*}{Excluded}
& Confusing (Table~\ref{tab:confusing-chars}) &
\bn{(ত,৩), (র,ব), (ড,ড়), (ঢ়,ঢ,ট), (য,য়), (০,0), (৪,8)} \\

& Rare / hard to type &
\bn{ঔ, ঋ, ঐ, ঊ, ৎ, ং, ঃ, ঁ, ঞ, ঙ, ঈ} \\

& Blends with noise &
\bn{্ (হসন্ত), ` (রেফ)} \\

& Invalid diacritics (Table~\ref{tab:vowel-problems}) &
\bn{ি, ী, ু, ূ, ৃ, ে, ৈ, ো, ৌ, ্র, ্য, া} \\

& Invalid conjuncts (Table~\ref{tab:conjunct-problems}) &
\bn{ক্ষ, ঙ্ক, ঙ্গ, জ্ঞ, ঞ্চ, ঞ্ছ, ঞ্জ, ত্ত, ষ্ণ, হ্ম, ণ্ড, \ldots} \\
\bottomrule
\end{tabular}
\caption{Selected and excluded Bengali characters for CAPTCHA generation}
\label{tab:used-removed-letters}
\end{table}

\begin{table}[h]
    \centering
    \setlength{\tabcolsep}{4pt}
    
    \begin{tabular}{c|c|c}
        \hline
        \textbf{Confusing Letters} & \textbf{Sample 1} & \textbf{Sample 2} \\ 
        \hline \hline
        
        (\bn{ড, ড়}) & 
        \parbox[c]{17mm}{\includegraphics[width=15mm, height=9mm]{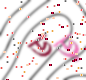}} & 
        \parbox[c]{17mm}{\includegraphics[width=15mm, height=9mm]{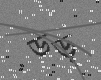}} \\  
        \hline
        
        (\bn{ঢ, ঢ়}) & 
        \parbox[c]{17mm}{\includegraphics[width=15mm, height=9mm]{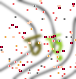}} & 
        \parbox[c]{17mm}{\includegraphics[width=15mm, height=9mm]{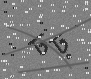}} \\  
        \hline

    \end{tabular}
    
    \caption{Examples of confusing letters in CAPTCHA}
    \label{tab:confusing-chars}
\end{table}

\begin{table}[h]
    \centering
    \setlength{\tabcolsep}{4pt}

    \begin{tabular}{c|c|c}
        \hline
        \textbf{\begin{tabular}[c]{@{}c@{}}Vowel Diacritics\end{tabular}} & 
        \textbf{\begin{tabular}[c]{@{}c@{}}Input Text\end{tabular}} & 
        \textbf{\begin{tabular}[c]{@{}c@{}}Transformation in Image\end{tabular}} \\ 
        \hline \hline
        
        \bn{"ো"} & \bn{গো} & 
        \parbox[c]{17mm}{\includegraphics[width=15mm, height=9mm]{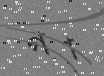}} \\ 
        \hline
        
        \bn{"ে"} & \bn{গেল} & 
        \parbox[c]{17mm}{\includegraphics[width=15mm, height=9mm]{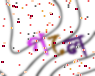}} \\ 
        \hline

    \end{tabular}
    \caption{Examples of Image Transformation Problem with Vowel Diacritics}
    \label{tab:vowel-problems}
\end{table}

\begin{table}[h]
    \centering
    \setlength{\tabcolsep}{4pt}

    \begin{tabular}{c|c|c}
        \hline
        \textbf{Conjuncts} & 
        \textbf{\begin{tabular}[c]{@{}c@{}}Input Text\end{tabular}} & 
        \textbf{\begin{tabular}[c]{@{}c@{}}Transformation in Image\end{tabular}} \\ 
        \hline \hline
        
        \bn{"ন" + "দ" = "ন্দ"} & \bn{মন্দ} & 
        \parbox[c]{17mm}{\includegraphics[width=15mm, height=9mm]{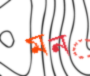}} \\ 
        \hline
        
        \bn{"ন" + "ন" = "ন্ন"} & \bn{অন্ন} & 
        \parbox[c]{17mm}{\includegraphics[width=15mm, height=9mm]{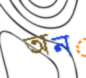}} \\ 
        \hline

    \end{tabular}
    \caption{Examples of Image Transformation Problem with Consonant Conjuncts}
    \label{tab:conjunct-problems}
\end{table}

\subsection{CAPTCHA Background}
\begin{figure}[h]
    \centering
    \newcommand{\subfigwidth}{0.28\columnwidth} 
    
    \begin{subfigure}{\subfigwidth}
        \centering
        \includegraphics[width=\linewidth]{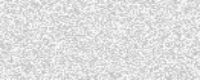}
        \caption{BG1: Variant 1}
        \label{background1}
    \end{subfigure}
    \hfil
    \begin{subfigure}{\subfigwidth}
        \centering
        \includegraphics[width=\linewidth]{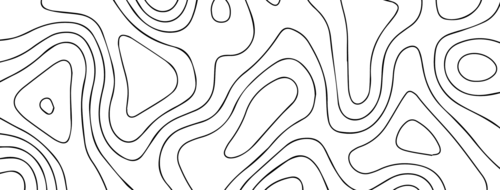}
        \caption{BG2: Variants 2 \& 3}
        \label{BGv2}
    \end{subfigure}
    
    \vspace{0.7em}
    
    \begin{subfigure}{\subfigwidth}
        \centering
        \includegraphics[width=\linewidth]{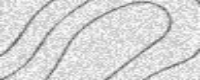}
        \caption{BG3: Variant 4}
        \label{BG4}
    \end{subfigure}
    \hfil
    \begin{subfigure}{\subfigwidth}
        \centering
        \includegraphics[width=0.85\linewidth]{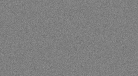}
        \caption{BG4: Variant 5}
        \label{RandomBVariant4}
    \end{subfigure}
    
    \vspace{0.7em}
    
    \begin{subfigure}{\subfigwidth}
        \centering
        \includegraphics[width=0.85\linewidth]{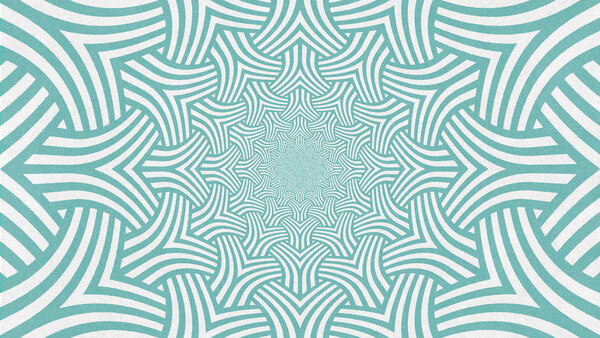}
        \caption{BG5: Variant 6}
        \label{RandomBVariant6}
    \end{subfigure}

    \caption{CAPTCHA Backgrounds.}
    \label{fig:cap_bg}
\end{figure}

\begin{figure}[h]
\centering
\begin{subfigure}{0.28\columnwidth}
\centering
    \includegraphics[width=\columnwidth]{img/Bg_images/BG2.png}
  \caption{Image for BG2}
  \label{background2}
\end{subfigure}
\begin{subfigure}{0.3\columnwidth}
\centering
    \includegraphics[width=0.5\columnwidth]{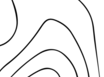}
  \caption{Random BG2 cropped part}
  \label{BG2Cropped}
\end{subfigure}%
~
\begin{subfigure}{0.3\columnwidth}
\centering
    \includegraphics[width=0.5\columnwidth]{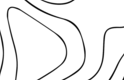}
  \caption{Random BG2 cropped part}
  \label{RandomBG2}
\end{subfigure}%
~
\begin{subfigure}{0.3\columnwidth}
    \includegraphics[width=0.5\columnwidth]{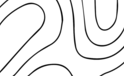}
  \caption{Random BG2 cropped part}
  \label{BG2CROPPED}
\end{subfigure}%

\begin{subfigure}{0.3\columnwidth}
    \includegraphics[width=\columnwidth]{img/Bg_images/BG3_1.png}
  \caption{Random BG3 cropped part}
  \label{RandomBG3Cropped}
\end{subfigure}%
~
\begin{subfigure}{0.3\columnwidth}
    \includegraphics[width=\columnwidth]{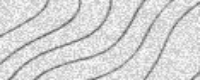}
  \caption{Random BG3 cropped part}
  \label{fig:sub6}
\end{subfigure}%
~
\begin{subfigure}{0.3\columnwidth}
    \includegraphics[width=\columnwidth]{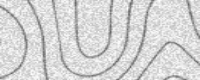}
  \caption{Random BG3 cropped part}
  \label{fig:sub7}
\end{subfigure}%

\begin{subfigure}{0.28\columnwidth}
      \centering
    \includegraphics[width=\columnwidth]{img/Bg_images/BG5.png}
  \caption{Image for BG5}
  \label{fig:sub8}
\end{subfigure}
\begin{subfigure}{0.3\columnwidth}
    \includegraphics[width=\columnwidth]{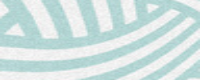}
  \caption{Random BG5 cropped part}
  \label{fig:sub9}
\end{subfigure}%
~
\begin{subfigure}{0.3\columnwidth}
    \includegraphics[width=\columnwidth]{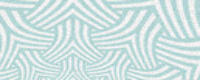}
  \caption{Random BG5 cropped part}
  \label{randomBackground}
\end{subfigure}%
~
\begin{subfigure}{0.3\columnwidth}
    \includegraphics[width=\columnwidth]{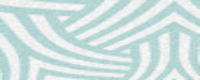}
  \caption{Random BG5 cropped part}
  \label{fig:sub11}
\end{subfigure}%

\caption{Random cropping of BG2, BG3, and BG5}
\label{fig:bg-cropping}
\end{figure}

In our scheme, we have built two CAPTCHA types with six variants, each with a unique combination of security and usability characteristics. We have used five different (BG1, BG2, BG3, BG4, BG5) background images in all six variants (Figure \ref{fig:cap_bg}). These are:
\begin{itemize}
    \item BG1 Background is for variant 1 CAPTCHA, a PIL-generated \cite{clark2015pillow} white noise image, and it is static for every CAPTCHA in Variant 1 (Figure \ref{fig:cap_bg} (a)).
    \item Next, BG2 Background is for variant 2 and variant 3 CAPTCHA, which is based on noisy structures as presented in \cite{vecteezyTopographicBackground}. Afterwards, we have cropped the image and constructed the CAPTCHA background dynamic, with each CAPTCHA having a different cropped part of the image for each sample of the variant (Figure \ref{fig:bg-cropping} (a)(b)(c)(d)).
    \item BG3 Background is for variant 4 CAPTCHA, a combination of BG1 and BG2. It is dynamic like BG2 (Figure \ref{fig:cap_bg} (c) and Figure \ref{fig:bg-cropping} (e)(f)(g)).
    \item BG4 (Figure \ref{fig:cap_bg} (d)) Background is for variant 5 CAPTCHA, another PIL-generated noisy static image like BG1.
    \item BG5 (Figure \ref{fig:cap_bg} (e)) Background is used in variant 6. Similar to BG3, this too is a combination of two different images, the first one, selected based on the noisy structures in it, is blended with BG1. We also applied the same cropping technique in this, making it a dynamic CAPTCHA like BG2 (Figure \ref{fig:bg-cropping} (i)(j)(k)).
\end{itemize}

\subsection{Use of Color}
\label{sec: use of color}
The use of color in text-based CAPTCHAs enhances resistance to automated attacks by increasing the difficulty of character segmentation and recognition. Random color variation disrupts color-based thresholding and segmentation techniques commonly used in automated CAPTCHA attacks \cite{kumar2022systematic}. At the same time, appropriate color variation can improve human readability, enabling users to solve CAPTCHAs more efficiently. Our approach is discussed below.
\begin{itemize}
    \item Except for BG5, all other backgrounds were black, white, or silver. Although this may seem weak, maintaining similar background and text colors is a reliable security feature against automated bot attacks \cite{shibbir2024evaluating}. We applied this criterion in BG1, BG4, and BG5, and omitted it in the other variants to maintain diversity.
    \item For variants 2, 3, and 4, we applied high-saturation, eye-catching colors on each character, randomly selected from a palette of RGB values spanning reds, oranges, yellows, greens, blues, purples, and dark tones. We deliberately avoided colors that could blend with the background or noise, as well as pale or low-saturation tones, to maintain human readability while preventing obscured letters \cite{Qin2011, kopp2017breaking}. The broad hue spread and per-letter randomization disrupt color-based clustering and segmentation algorithms commonly used by automated CAPTCHA solvers, while high contrast and luminance variability preserve usability, achieving an effective balance between security and legibility \cite{fortunecs229, zhu2010attacks}. Following is the color list in RGB: (248, 112, 187), (74, 80, 171), (137, 83, 5), (237, 88, 78), (1, 85, 20), (71, 215, 108), (222, 46, 60), (25, 67, 222),(113, 88, 74), (220, 31, 243),(140, 15, 5), (168, 16, 42), (250, 70, 7), (115, 90, 30), (115, 90, 30), (16, 9, 22), (19, 4, 155), (101, 93, 3), (67, 10, 103), (247, 133, 20), (49, 81, 224), (139, 208, 9), (134, 1, 39),
\end{itemize}

\subsection{Morphing of Letters}
Morphing purposefully warps or distorts the characters in a CAPTCHA such that they are still identifiable to humans but more difficult for computers to decipher. Characters can be stretched, bent, twisted, blurred, or warped as part of morphing, among other changes. With each CAPTCHA, a different modification may be applied; it may be randomized or decided by a particular algorithm.

We have applied morphing in variants 1, 2, and 4 (Table \ref{tab:morphing-rotation-chars}). Variants 1 and 4 have the same kind of morphing which is stretched, bent, and blurred, however, variant 2 mainly uses warping which we have classified as CAPTCHA rotation and discussed in Section \ref{subsec:rotation}.

\begin{table*}[h]
\centering
\setlength{\tabcolsep}{3pt}
\renewcommand{\arraystretch}{1.1}

\begin{tabular*}{\linewidth}{@{\extracolsep{\fill}} c c c c}
\toprule
\textbf{Before Morphing} & \textbf{After Morphing} & \textbf{Before Rotation} & \textbf{After Rotation}\\
\midrule \midrule
\parbox[c]{4em}{\includegraphics[width=11mm, height=7mm]{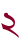}} &
\parbox[c]{4em}{\includegraphics[width=11mm, height=7mm]{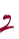}} &
\parbox[c]{4em}{\includegraphics[width=11mm, height=7mm]{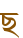}} &
\parbox[c]{4em}{\includegraphics[width=11mm, height=7mm]{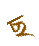}} \\

\parbox[c]{4em}{\includegraphics[width=13mm, height=8mm]{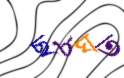}} &
\parbox[c]{4em}{\includegraphics[width=13mm, height=8mm]{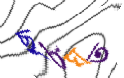}} &
\parbox[c]{4em}{\includegraphics[width=13mm, height=8mm]{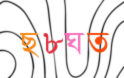}} &
\parbox[c]{4em}{\includegraphics[width=13mm, height=8mm]{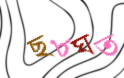}} \\
\bottomrule
\end{tabular*}

\caption{Examples of morphing and rotation transformations of letters in CAPTCHA.}
\label{tab:morphing-rotation-chars}
\end{table*}

\subsection{Rotation}
\label{subsec:rotation}
The characters in text CAPTCHAs are are purposefully rotated so that they are still identifiable to humans but more difficult for computers to decipher \cite{boyter2015, brodic2019captcha}. Rotation can take many different forms, such as rotating the characters at a certain angle or randomly choosing each characters rotational orientation. With each CAPTCHA, a different modification may be applied; it may be randomized or decided by a particular algorithm. Also, since the letters may appear in unanticipated orientations, the rotation might make it more difficult for attackers to create templates or models for recognizing CAPTCHAs.

\begin{figure}[h]
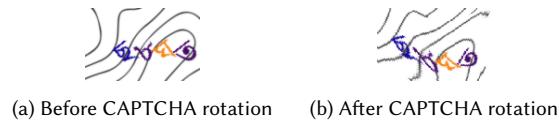

\centering
\begin{subfigure}{0.25\textwidth}
\centering
    \includegraphics[width=0.4\textwidth]{img/Morphing/befor_morph.png}
  \caption{Before CAPTCHA rotation}
  \label{beforeRotation}
\end{subfigure}%
~
\begin{subfigure}{0.25\textwidth}
\centering
    \includegraphics[width=0.4\textwidth]{img/Morphing/after_morph.png}
  \caption{After CAPTCHA rotation}
  \label{CAPTCHA_rotation}
\end{subfigure}%

\caption{Effects of CAPTCHA rotation}
\label{fig:cap-rotation}
\end{figure}

\paragraph{\textbf{CAPTCHA Rotation:}} In CAPTCHA variant 2, we have used CAPTCHA rotation (Figure \ref{fig:cap-rotation}) meaning we have rotated every content in that particular CAPTCHA with the same amount. In this case, it is performed using sine and cosine curves which causes a wave-like rotation of the background. It skews the background and creates some more room. Also, we can allow the newly opened region to be automatically assigned a color, in this case, white.

\paragraph{\textbf{Letter Rotation:}} In CAPTCHA variant 2, 3, and 5, we have applied letter rotation while pasting that letter on the background.

\subsection{Other Noises}
Adding noises to text CAPTCHAs is a common technique used to make them more difficult for automated programs to solve. In a noisy CAPTCHA, additional elements are added to the image to make it harder for a computer to accurately segment the characters from the background \cite{boyter2015, brodic2019captcha}.

To increase the difficulty and effectiveness of CAPTCHAs, we incorporated various types of noise which are discussed below:
\begin{itemize}
    \item \textbf{Line noise:} The characters are disguised by adding arbitrary curves or lines to the image background.
    \item \textbf{Dot noise:} To add visual clutter, arbitrary dots or speckles are added to the image.
    \item \textbf{Distortion noise:} The characters are slightly distorted or warped, making it harder for a computer to recognize them.
    \item \textbf{Blurriness:} Blurring a text CAPTCHA can make it more difficult for bots. However, the amount of additional security this provides can vary depending on how the blurring is done.
\end{itemize}

Although noise might be useful in preventing automated assaults, it is important to keep the CAPTCHA readable and friendly for human users. Certain users, especially those with visual impairments or cognitive problems, may find it challenging to read the characters when there is too much background noise. Because of this, it is crucial to test CAPTCHAs with a variety of users and ensure that it stays usable by everyone while providing enough protection against automated assaults.

Depending on the specific implementation and intended level of difficulty, several types and levels of noise may be employed as discussed next:

\subsubsection{Lines}
We have used two types of lines in CAPTCHA variants 1, 4, 5, and 6. One is just some general noisy lines around the text, and the other one is called a Hollow scheme, noisy line(s) that go through the text. Line noise was applied only to selected variants to preserve diversity among the six CAPTCHA designs and to evaluate its effect on both security and usability. If all variants contained the same line-based distortion, it would be difficult to isolate the contribution of this feature to recognition difficulty and user performance.

\subsubsection{Dotted Like Shape}
\begin{figure}[h]
\centering
\begin{subfigure}{0.13\columnwidth}
    \includegraphics[width=0.5\columnwidth]{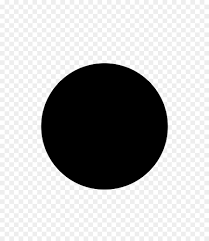}
  
  \label{dottedNoise}
\end{subfigure}%
~
\begin{subfigure}{0.13\columnwidth}
    \includegraphics[width=0.5\columnwidth]{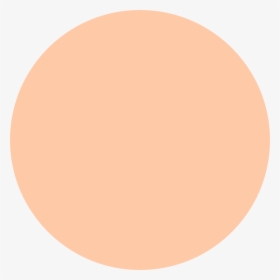}
 
  \label{DottedNoise}
\end{subfigure}%
~
\begin{subfigure}{0.13\columnwidth}
    \includegraphics[width=0.5\columnwidth]{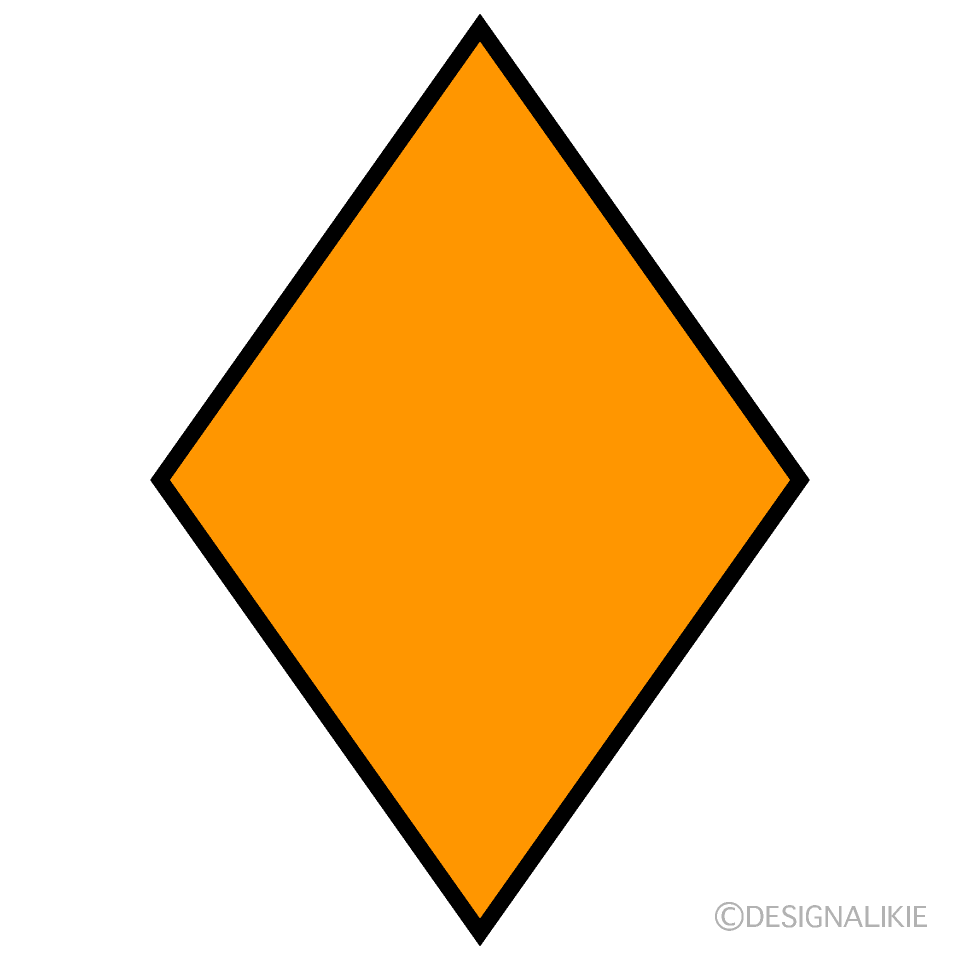}
  
  \label{dottedYello}
\end{subfigure}%

\begin{subfigure}{0.13\columnwidth}
    \includegraphics[width=0.5\columnwidth]{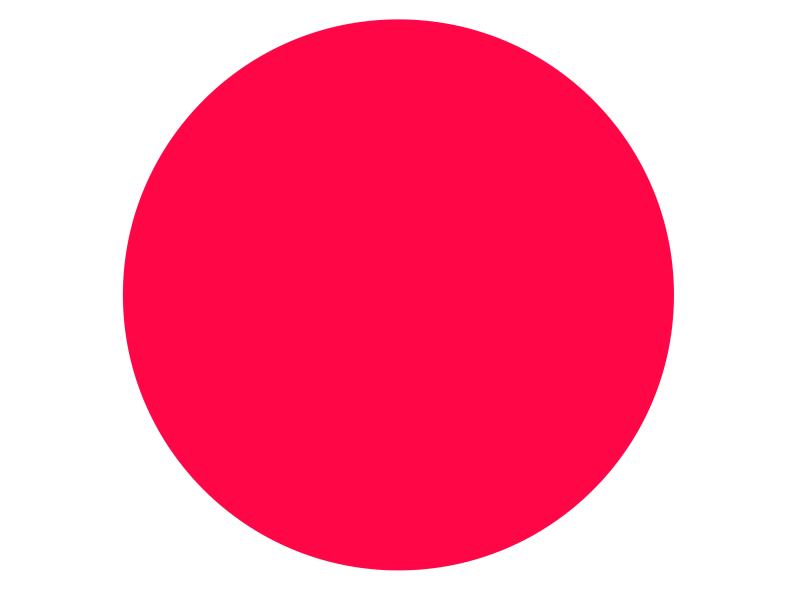}
  \label{pinkCircle}
\end{subfigure}%
~
\begin{subfigure}{0.13\columnwidth}
    \includegraphics[width=0.5\columnwidth]{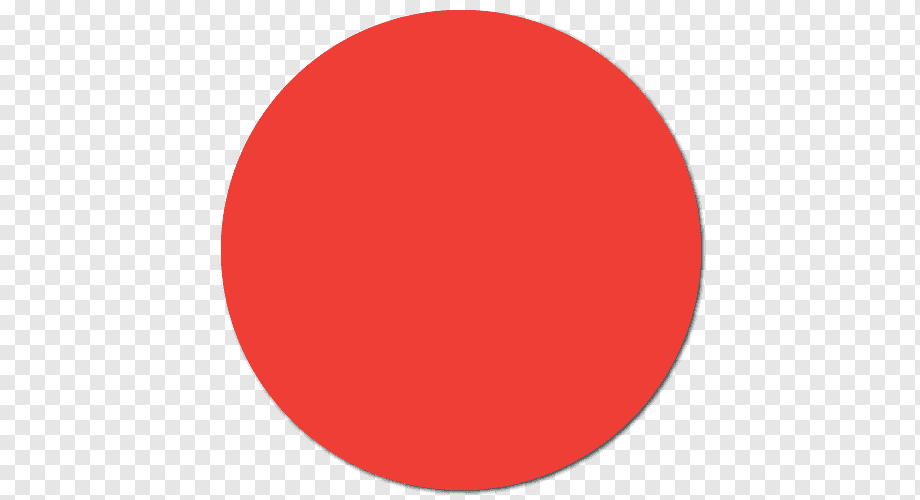}

  \label{redCircle}
\end{subfigure}%
~
\begin{subfigure}{0.13\columnwidth}
    \includegraphics[width=0.5\columnwidth]{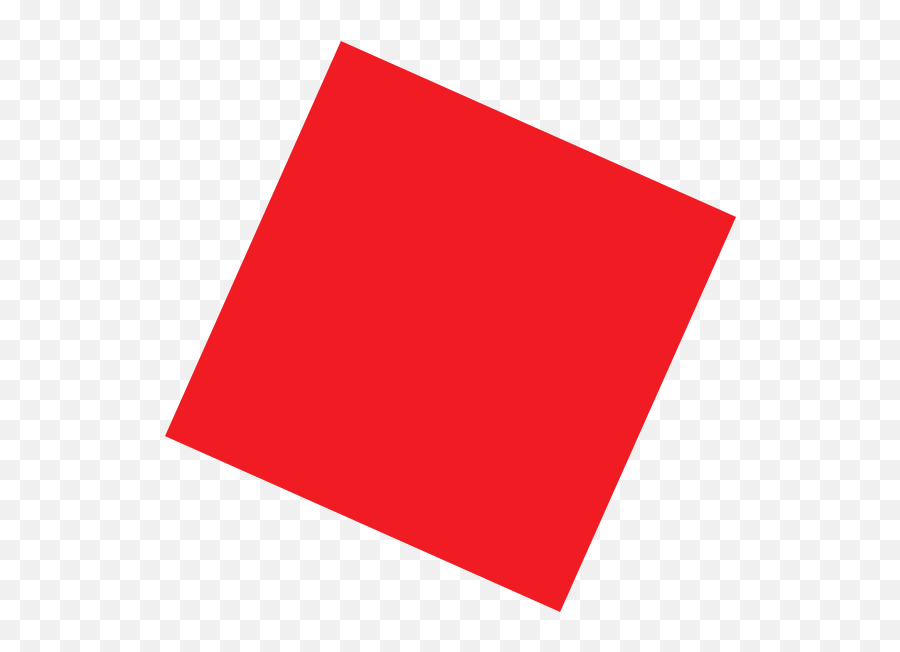}

  \label{rotatedSquare}
\end{subfigure}%
~
\begin{subfigure}{0.13\columnwidth}
    \includegraphics[width=0.5\columnwidth]{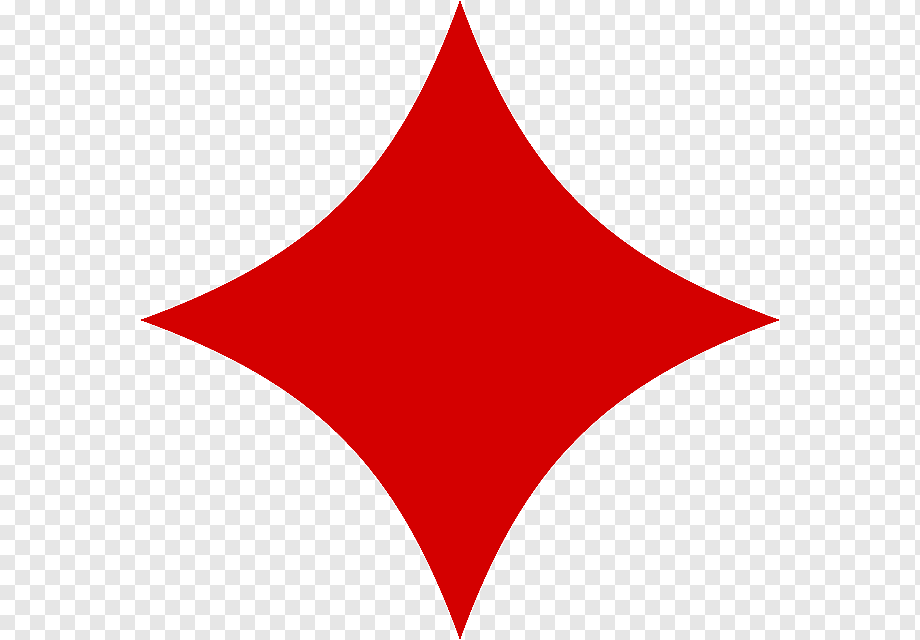}

  \label{star}
\end{subfigure}%

\caption{Dotted Shape Noises}
\label{fig:dotted-noise}
\end{figure}

\begin{figure}[h]
\centering
\begin{subfigure}{0.28\columnwidth}
\centering
    \includegraphics[width=0.8\columnwidth, height=0.3\columnwidth]{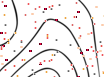}
  
  \label{dottedShape}
\end{subfigure}%
~
\begin{subfigure}{0.28\columnwidth}
\centering
    \includegraphics[width=0.8\columnwidth, height=0.3\columnwidth]{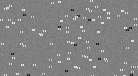}
  
  \label{dottedShape2}
\end{subfigure}%
~

\caption{Dotted shape noise on Background}
\label{fig:dotted-bg}
\end{figure}

In CAPTCHA variants 3 and 5, we employed noise in the form of dotted shapes.This noise was applied selectively to avoid excessive visual clutter and to maintain readability across variants. Several rotations of various shapes, including circles, squares, and diamonds, were employed. Then, we reduced their size to that of a dot and randomly scattered about 100 of them over the text image, selecting from a variety of shapes automatically. Figure \ref{fig:dotted-noise} presents the type of noise adopted in our scheme and their effects on the CAPTCHA image is illustrated in Figure \ref{fig:dotted-bg}.

\subsubsection{Blurriness}
\begin{figure}[h]
\centering
\begin{subfigure}{0.35\columnwidth}
    
    \centering
    \includegraphics[width=0.4\columnwidth]{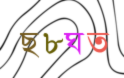}
  \caption{Before gaussian blur}
  \label{fig:gausBlurBef}
\end{subfigure}%
~
\begin{subfigure}{0.35\columnwidth}
\centering
    \includegraphics[width=0.45\columnwidth]{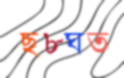}
 \caption{After gaussian blur}
  \label{fig:gausBlurAft}
\end{subfigure}%

\begin{subfigure}{0.35\columnwidth}
\centering
    \includegraphics[width=0.45\columnwidth]{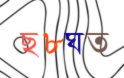}
  \caption{Before morphing blur}
  \label{fig:morphBlurBef}
\end{subfigure}%
~
\begin{subfigure}{0.35\columnwidth}
\centering
    \includegraphics[width=0.4\columnwidth]{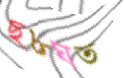}
 \caption{After morphing blur}
  \label{fig:morphBlurAft}
\end{subfigure}%

\caption{Different Blurring Effects}
\label{fig:blur-effects}
\end{figure}

We applied a filter module from the PIL package named ImageFilter along with the GaussianBlur method  \cite{PIL} in CAPTCHA variants 2, 3, and 5. We used radius, a standard deviation of the Gaussian kernel, a value ranging from 0.5 to 1.2 (Figure \ref{fig:gausBlurBef} and Figure \ref{fig:gausBlurAft}).

We used letter morphing in CAPTCHA variants 1, 4, and 6, which also caused the texts to become blurry, as presented in Figure \ref{fig:morphBlurBef} and Figure \ref{fig:morphBlurAft}.

\subsubsection{Distortion}
The distortion or morphing process primarily involves making letters and/or numbers that are distorted, misaligned, or uneven in the text. This is one of the anti-recognition techniques employed to make image processing more difficult. Here, we used letter morphing in CAPTCHA variants 1, 4, and 6, which also caused the texts to become blurry, as presented in Figure \ref{fig:morphBlurBef} and Figure \ref{fig:morphBlurAft}.

\section{Developing CAPTCHA Variants}
\label{section5: captcha variants design}
We have developed the Bengali CAPTCHA having six variants with different sets of security features as discussed in Section \ref{section4: design of a secure and usable bengali cAPTCHA}. The six variants were developed as an initial exploratory set of Bengali text-CAPTCHA prototypes, not as an exhaustive exploration of all possible designs. We selected six variants to balance design diversity, Bengali-script rendering constraints, and participant workload, since each participant solved three samples per variant. The variants were grouped into two broad families: CAP\_1 variants use rotated characters and Gaussian blur, while CAP\_2 variants use letter morphing, morphing-induced blur, and multi-structure text. Therefore, the six variants should be interpreted as representative exploratory prototypes for comparing major Bengali CAPTCHA design choices rather than as a complete coverage of the entire design space. Table \ref{tab:captcha-features} lists the security features employed in each variant. In the following, we discus different implementation aspects of our proposed Bengali CAPTCHA. 

\begin{table*}[!htb]
\centering
\small 
\setlength{\tabcolsep}{4pt}
\renewcommand{\arraystretch}{1.1}
\caption{Security features employed in different CAPTCHA types and their variants}
\label{tab:captcha-features}
\begin{tabular}{c p{0.22\linewidth} c p{0.60\linewidth}} 
\toprule
\textbf{Type} & \textbf{Common Features} & \textbf{Var.} & \textbf{Variant Creation} \\
\midrule \midrule
\multirow{3}{*}{CAP\_1} 
& \multirow{3}{=}{No letter morphing; rotated letters; single font; Gaussian blur} 
& V2 & Dynamic BG2; blur (0.5--0.8); char. rot. (11--30$^\circ$); CAPTCHA curve rot.; multi-color text \\
& & V3 & Dynamic BG2; blur (0.8--1.2); char. rot. (18--45$^\circ$); multi-color text; dotted shapes \\
& & V5 & Static BG4; blur (0.5--0.8); char. rot. (18--45$^\circ$); hollow lines (3--5); fixed color \\
\midrule
\multirow{3}{*}{CAP\_2} 
& \multirow{3}{=}{Letter morphing blur; multi-structure text; multiple fonts} 
& V1 & Static BG1; letter morphing; hollow lines (3--5); morphing blur \\
& & V4 & Dynamic BG3; hollow lines (3--5); multi-color text; letter morphing; morphing blur \\
& & V6 & Dynamic BG5; similar BG; text color; hollow lines (3--5); letter morphing; morphing blur \\
\bottomrule
\end{tabular}
\end{table*}

\begin{table*}[!h]
    \footnotesize
    \begin{tabular*}{\linewidth}{@{\extracolsep{\fill}}c c c c c c}
        \toprule
        \textbf{Variants} & \textbf{Example} & \textbf{Anti-Preprocess} & \textbf{Anti-Segmentation} & \textbf{Anti-Recognition} & \textbf{Vulnerabilities} \\
        \midrule \midrule
        
        V1 & \parbox[c]{3em}{\includegraphics[width=12mm, height=8mm]{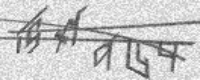}} & 
        \begin{tabular}[c]{@{}c@{}}Textured Background,\\ Noisy Background,\\Same colored\\ Character and Background,\\ Rotation, Distortion \\Multi-Struct, Warping \end{tabular} &  
        Background Blurring & 
        Rotation, Distortion, Multi-Struct & 
        \begin{tabular}[c]{@{}c@{}} Constant Font\\ Binary Background \end{tabular} \\ \hline

        V2 & \parbox[c]{3em}{\includegraphics[width=12mm, height=8mm]{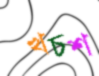}} & 
        \begin{tabular}[c]{@{}c@{}}Textured Background,\\ Random Background,\\ Rotation, Distortion,\\Multi-Colored Text, \\ Warping \end{tabular} &  
        \begin{tabular}[c]{@{}c@{}}Background Blurring,\\Overlapping Chars,\\ Deformation \end{tabular} & 
        Rotation, Distortion & 
        \begin{tabular}[c]{@{}c@{}} Constant Font \end{tabular} \\ \hline

        V3 & \parbox[c]{3em}{\includegraphics[width=12mm, height=8mm]{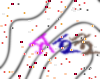}} & 
        \begin{tabular}[c]{@{}c@{}}Textured Background,\\ Noisy Background,\\Random Background,\\ Rotation, Distortion,\\Multi-Colored Text, \\Warping \end{tabular} &  
        \begin{tabular}[c]{@{}c@{}}Background Blurring,\\Overlapping Chars,\\ Deformation \end{tabular} & 
        Rotation, Distortion & 
        \begin{tabular}[c]{@{}c@{}} Constant Font \end{tabular} \\ \hline

        V4 & \parbox[c]{3em}{\includegraphics[width=12mm, height=8mm]{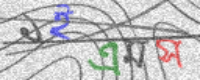}} & 
        \begin{tabular}[c]{@{}c@{}}Textured Background,\\ Noisy Background,\\ Random Background,\\ Rotation, Distortion,\\Multi-Colored Text, \\Multi-Struct, Warping \end{tabular} &  
        Background Blurring & 
        Rotation, Distortion, Multi-Struct & 
        \begin{tabular}[c]{@{}c@{}} - \end{tabular} \\ \hline

        V5 & \parbox[c]{3em}{\includegraphics[width=12mm, height=8mm]{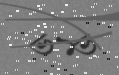}} & 
        \begin{tabular}[c]{@{}c@{}}Textured Background,\\ Noisy Background,\\Random Background,\\Same Colored \\ Character and Background,\\ Rotation, Distortion \\Warping \end{tabular} & 
        \begin{tabular}[c]{@{}c@{}}Background Blurring,\\Overlapping Chars,\\ Deformation \end{tabular} & 
        Rotation, Distortion & 
        \begin{tabular}[c]{@{}c@{}} Constant Font \end{tabular} \\ \hline

        V6 & \parbox[c]{3em}{\includegraphics[width=12mm, height=8mm]{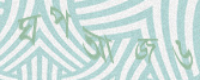}} & 
        \begin{tabular}[c]{@{}c@{}}Textured Background,\\ Noisy Background,\\Random Background,\\Same Colored \\ Character and Background,\\ Rotation, Distortion \\Multi-Struct, Warping \end{tabular} &  
        \begin{tabular}[c]{@{}c@{}}Background Blurring,\\ Deformation \end{tabular} & 
        Rotation, Distortion, Multi-Struct & 
        \begin{tabular}[c]{@{}c@{}} - \end{tabular} \\ 

        \bottomrule
    \end{tabular*}
    \caption{Security and vulnerability features of the Bengali CAPTCHAs}\label{table:BengaliSecurityFeature}
\end{table*}

We implemented the Bengali CAPTCHA using the Pillow \cite{clark2015pillow} and OpenCV \cite{opencv_py} libraries. and by extending the codebase of an open-source Github repository \cite{Kuszaj_2017}. We used the implementation structure of the codebase as a starting point rather than coding entirely from the scratch. Then, we extended it to incorporate our security principles, and Bengali language specific design requirements. We did this way to speed up the implementation with a stable codebase, so we could focus on our proposed new security features. Now, let us elaborate on the implementation procedures in detail.

\begin{enumerate}[label=(\roman*)]

    \item \textbf{Generate a random text string:} 
    We used the Python random module to produce a random Bengali text string.

    \item \textbf{Generate a blank image:} 
    For each CAPTCHA, we began by generating a blank image using the Pillow module. The image dimensions were selected to ensure that the entire text string could be accommodated without truncation. Once the base image was created, the characters were rendered onto it, followed by the application of noise, distortions, and other visual effects.

    \item \textbf{Use a background image:} 
    For CAPTCHA variants 1 and 5, we used two static background images, BG1 and BG4 (Figure~\ref{fig:cap_bg}). These images were resized according to the blank image size and applied directly as the background. For the other variants (2, 3, 4, and 6), a random window from the background image was selected each time a CAPTCHA was generated (Figure \ref{fig:bg-cropping}).

    \item \textbf{Draw the text string:} 
    The characters from the randomly generated text string were drawn onto the blank image. The font type and 
    size were adjusted according to the design specifications.

    \item \textbf{Random blur addition to the CAPTCHA:} 
    One technique to add random blur to the CAPTCHA is to use a filter from the Python image package Pillow. Random blur was applied using the GaussianBlur method from Python Pillow’s ImageFilter module. The blur radius was randomly selected within the range specified for each variant.

    \item \textbf{Random distortion addition to the CAPTCHA:} 
    We applied several types of distortions based on the following design criteria.  

    \begin{enumerate}
        \item \textbf{Transformation:} 
        We added three kinds of transformations to our CAPTCHA: background and line transformation and character morphing (with values ranging approximately from 0.2 to 0.8).

        \item \textbf{Rotation:} 
        We rotated the characters in some CAPTCHAs by a random angle between 11 and 45 degrees. Different angle ranges were tested, and this range provided a good balance between security and usability~\cite{gossweiler2009s}.
    \end{enumerate}

    \item \textbf{Random lines addition to the CAPTCHA:} 
    Random lines were added to make it harder for automated programs to detect the characters.

    \item \textbf{Use of color:} 
    Images are represented in Pillow as a two-dimensional grid of pixels. Each pixel can have a color represented by a tuple of red, green, and blue (RGB) values ranging from 0 to 255. Pillow also supports other color modes, such as 
    grayscale and CMYK. We can adjust brightness, contrast, and color balance to refine the CAPTCHA appearance.

\end{enumerate}
We note that not all security features used in the variants are unique to Bengali script. Features such as blur, rotation, background noise, and color variation are common in text-CAPTCHA design. The Bengali-specific contribution lies in adapting these mechanisms to Bengali script constraints, including character similarity, Bengali input methods, font rendering behavior, exclusion of problematic diacritics and conjuncts, and the usability-security trade-off for Bengali-speaking users.

\section{Security Evaluation}
\label{securityAnlaysis}
This section discusses the security aspects of the Bengali CAPTCHA we developed by demonstrating its level of security with the aid of an Evaluation Framework presented in \cite{shibbir2024evaluating}.

\subsection{Evaluation Framework}
In \cite{shibbir2024evaluating}, the authors devised an evaluation framework based on two aspects, namely: Security and Vulnerability features. Both security features (anti-preprocessing, anti-segmentation, and anti-classification) and vulnerability features (pre-processing, segmentation, and classification) can be understood across three stages. The preprocessing stage refers to initial image handling, such as noise reduction, binarization, or smoothing, which aims to simplify attacks or resist them. The segmentation stage involves separating characters from the background or from each other, where robust CAPTCHAs seek to make such separation difficult. The classification stage involves the recognition and labeling of characters, typically using machine learning models, where security mechanisms introduce distortions and obfuscations to hinder accurate classification.

In this work, we utilized the evaluation framework to assess the security and vulnerability features (Table \ref{table:BengaliSecurityFeature}) of the developed Bengali CAPTCHA. Using the evaluation framework, we can predict the security level of the CAPTCHA variations using two equations:

\begin{equation} \label{circle_area_eq}
\Delta D_{C_i} = \sum (SF_C) - \sum (VF_C)
\end{equation}

\begin{equation}
\label{eq:label}
  C_i =
    \begin{cases}
      \text{Vulnerable} & \Delta D_{C_i} \leq 0\\
      \text{Moderately Secure} & 0 < \Delta D_{C_i} \leq 6\\
      \text{Secure} & \Delta D_{C_i} > 6
    \end{cases}       
\end{equation}

Equation \ref{circle_area_eq} represents the difference (denoted with $\Delta D$) between the total number of security features (denoted as $SF$) and vulnerability features (denoted as $VF$) for a particular CAPTCHA $C$ belonging to a variation $i$. We can divide all text CAPTCHAs into three categories, as shown in Equation \ref{eq:label}, based on the numerical differences between security features and vulnerability features. The threshold values of the Vulnerable, Moderately Secure and Secure in Equation \ref{eq:label} are adopted from the work of CAPTCHA evaluation framework \cite{shibbir2024evaluating} used in this study. These thresholds provide a heuristic categorization based on the difference between security and vulnerability features. We acknowledge that this framework-based classification is not a substitute for attack-based validation; therefore, we further evaluated the CAPTCHA variants using reak world attacks.

Based on this evaluation, we predicted the security levels of our CAPTCHA. In Table \ref{table:ourAssumtionBengaliCaptcha}, we present the difference between the security and vulnerability features of each of our developed Bengali CAPTCHA variants and the predicted security level associated with it. We can see that all the CAPTCHA variants are predicted as Secure.
\begin{table}[h]
\centering
\begin{tabular}{c c c c c}
\toprule
\textbf{V} & \textbf{SF} & \textbf{VF} & \textbf{D} & \textbf{SL}\\
\midrule \midrule
V1 & 11 & 2 & 9  & Secure \\
V2 & 11 & 1 & 10 & Secure \\
V3 & 12 & 1 & 11 & Secure \\
V4 & 12 & 0 & 12 & Secure \\
V5 & 12 & 1 & 11 & Secure \\
V6 & 13 & 0 & 13 & Secure \\
\bottomrule
\end{tabular}

\caption{Our assumptions on the level of security of Bengali CAPTCHAs based on the Evaluation Framework. Here V = Variants, SF = Security Features, VF = Vulnerability Features, D = Differences, SL = Security Level.}
\label{table:ourAssumtionBengaliCaptcha}
\end{table}
\subsection{Threat Model}

The real world attack in this work considers an automated adversary attempting to solve the proposed Bengali CAPTCHA without human assistance. The adversary is assumed to have access to the CAPTCHA image and may apply standard image-processing operations such as binarization, blurring, erosion, dilation, noise removal, segmentation, OCR, and post-processing. The adversary may also use publicly available or pre-trained Bengali OCR models.

The adversary is not assumed to have access to the server-side CAPTCHA solution, the random generation seed, or the internal challenge-generation parameters at runtime. We also do not consider attacks based on compromising the server, stealing session tokens, bypassing the web application logic, or using human CAPTCHA-solving farms. These attacks are important in practice but are outside the scope of a text-recognition-based CAPTCHA evaluation.

\subsection{Security Performance}
To validate the security of the proposed Text-Based Bengali CAPTCHA against traditional bots, We have generated 1000 samples for our each CAPTCHA variants (6000 in total) and performed attack on them by the configurations of this work \cite{shibbir2024evaluating}. This attack employs a standard three-stage approach consisting of pre-processing, optical character recognition (OCR), and post-processing. In the pre-processing stage, multiple noise-removal and enhancement techniques were applied to isolate character information. With the processed images, we assessed 8 pre-trained recognition models for Bengali language \cite{tesseract-ocr-ben,tesseract-ocr-tableBenTrain,tesseractIndic, PyBnDataset1,githubTesseractBen, vintasoft,tesseract-ocr-ben-best, Tesseract-Ocr_fast} to identify the optimal configuration for Bengali text recognition. Despite exhaustive pre-processing attempts, the embedded security features prevented effective character segmentation. Consequently, OCR failed to produce meaningful recognition results across all models. Post-processing techniques, which rely on identifiable patterns in recognized text, were therefore infeasible. This attack-based evaluation confirms that the developed CAPTCHA scheme remains resistant under realistic automated attack conditions.

Table \ref{table:attackBengali} presents the results of our security analysis. The average recognition rate is defined as the mean percentage of characters correctly identified by the attack algorithm. A higher average prediction rate indicates a more effective attack and, consequently, weaker CAPTCHA robustness. It is clear from the table that not a single full CAPTCHA was broken by the attack. The highest average recognition rate observed for the proposed CAPTCHA mechanism was 19.86\% (Variant 6) by this pretrained model \cite{tesseractIndic}. The maximum average recognition rates for all variants can be seen in Table \ref{table:attackBengali}. Across 6,000 CAPTCHA samples, two pretrained models \cite{vintasoft, tesseract-ocr-best} were able to fully recognize only one sample. Furthermore, since no variation yielded a recognition rate exceeding 20\% (see Table \ref{table:attackBengali}), we can conclude that all developed CAPTCHA variants demonstrate significant robustness against the tested attacks. Additional information about the attack can be found in the Section \ref{sec:attackBangladeshiCAPTCHA}.

\subsection{Preliminary Probe Against Vision-Language Models}
Recent vision-language models (VLMs)~\cite{radford2021learning} and browser-based AI agents introduce an attack vector not captured by the OCR-centric pipeline~\ref{sec:attackBangladeshiCAPTCHA}. To estimate exposure, a preliminary probe was conducted using the Comet browser agent~\cite{comet} against samples from each of the six variants. The agent was prompted with the natural-language instruction to read the Bengali text in the displayed image. Across the probed samples, no full CAPTCHA was correctly solved. Owing to the rapid pace of VLM development and the broad space of prompting strategies, this probe is reported as preliminary evidence rather than a formal security claim. A systematic evaluation against frontier multimodal models under controlled prompting protocols is identified as a primary direction for future work, and is necessary before the proposed scheme can be claimed to resist a fully modern adversary.

\begin{table*}[t] 
    \centering
    \caption{Attack results on Bengali Text-based CAPTCHAs}\label{table:attackBengali}
    \begin{tabular*}{\linewidth}{@{\extracolsep{\fill}}cccccc}
        \toprule
        \textbf{Variation} & \textbf{Example} & \textbf{Preprocessed} & \textbf{\shortstack{Total\\Characters}} & \textbf{\shortstack{Recognized\\Characters}} & \textbf{\shortstack{Average\\Recognition Rate}} \\
        \midrule \midrule
         1 & \parbox[c]{4em}{\includegraphics[width=15mm, height=9mm]{img/captcha_variants/captcha1.png}} & \parbox[c]{4em}{\includegraphics[width=15mm, height=9mm]{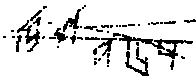}} & 3985 & 659 & 16.54\% \\ 
         2 & \parbox[c]{4em}{\includegraphics[width=15mm, height=9mm]{img/captcha_variants/captcha2.png}} & \parbox[c]{4em}{\includegraphics[width=15mm, height=9mm]{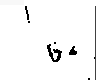}} & 3482 & 195 & 5.60\%  \\ 
         3 & \parbox[c]{4em}{\includegraphics[width=15mm, height=9mm]{img/captcha_variants/captcha3.png}} & \parbox[c]{4em}{\includegraphics[width=15mm, height=9mm]{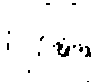}} & 3472 & 2   & 0.06\%  \\ 
         4 & \parbox[c]{4em}{\includegraphics[width=15mm, height=9mm]{img/captcha_variants/captcha4.png}} & \parbox[c]{4em}{\includegraphics[width=15mm, height=9mm]{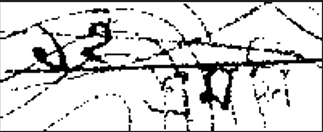}} & 3984 & 3   & 0.08\%  \\ 
         5 & \parbox[c]{4em}{\includegraphics[width=15mm, height=9mm]{img/captcha_variants/captcha5.png}} & \parbox[c]{4em}{\includegraphics[width=15mm, height=9mm]{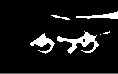}} & 3477 & 271 & 7.79\%  \\ 
         6 & \parbox[c]{4em}{\includegraphics[width=15mm, height=9mm]{img/captcha_variants/captcha6.png}} & \parbox[c]{4em}{\includegraphics[width=15mm, height=9mm]{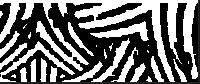}} & 4021 & 799 & 19.87\% \\ 
        \bottomrule
    \end{tabular*}
\end{table*}

\begin{table*}[!htb]
\centering
\small
\setlength{\tabcolsep}{3pt}
\caption{Comparison of security and vulnerability features across CAPTCHA processing stages}
\label{tab:security_vs_vulnerability}
\begin{tabular}{p{0.42\linewidth} p{0.26\linewidth} p{0.26\linewidth}}
\toprule
\textbf{Pre-process} & \textbf{Segmentation} & \textbf{Classification} \\
\midrule \midrule
\multicolumn{3}{l}{\textbf{Security Features}} \\
\midrule
Textured, noisy, random background; color variation; multi-layer structure; hollow scheme; rotation; warping; multi-colored text 
& Overlapping characters; connected characters; deformation; blurring
& Rotation; wrapping; distortion \\
\midrule
\multicolumn{3}{l}{\textbf{Vulnerability Features}} \\
\midrule
Constant background; binary color 
& Aligned characters 
& Letters/numbers only; fixed case; constant font; dictionary words \\
\bottomrule
\end{tabular}
\end{table*}

\section{Usability Evaluation}
\label{sec:UsabilityEvaluation}
Real human users may have trouble reading and solving CAPTCHAs if complex security features are employed in a CAPTCHA. Therefore, to test the ease and effectiveness with which users can complete any newly proposed CAPTCHA, it is important to test the usability of CAPTCHA. Hence, we conducted a usability evaluation of our proposed CAPTCHA which is presented in this section.

\subsection{Usability Study Methodology}

\begin{figure}[h]
\centering
\includegraphics[width=0.5\linewidth]{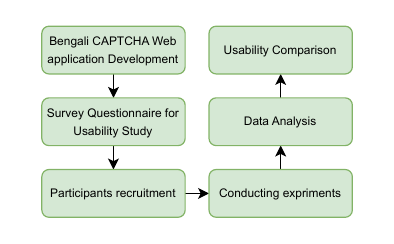}
\caption{Methods to follow for Usability Study}
\label{usabilityStudyFig}
\end{figure}

As per \cite{nielsen1994usability}, any user-centred interaction design relies on usability testing to evaluate products from a user’s perspective. This also applies to the newly proposed CAPTCHA scheme. The methods we followed during our usability study are presented in Figure \ref{usabilityStudyFig}.

\subsection{Web application deploying the Bengali CAPTCHA scheme}
A web application was developed in which the proposed CAPTCHA scheme was deployed. The application serves two main purposes. Firstly, we wanted to test the practicality of our CAPTCHA in a real-world application. Secondly, we used the web application for conducting usability testing to evaluate the usability of our Bengali CAPTCHA. The application provides a real-world experience for users, recording the time they take to solve each CAPTCHA sample. The web application was deployed online within a public domain \footnote{\url{https://captcha.pythonanywhere.com/}}. 

The application is compatible with any device with an internet connection, including PCs, smartphones, tablets, and more. The participant used their own device to access the web application. Using the participants' own devices has a great advantage in usability measures. As the user uses his/her device daily, our CAPTCHA will get an accurate, user-friendly judgment on the own device of a user. Also, different users possess different devices. Therefore, it will be tested on the maximum number of devices.

The application dynamically generates CAPTCHAs with various levels of randomness to enhance security. These features include random numbers of characters, lines, hollow schemes, background textures, morphing, and rotation. Consequently, every time a user encounters a CAPTCHA, a new one is generated dynamically on the web application’s backend. The application also records whether the user successfully solves the CAPTCHA. After solving three CAPTCHA samples successfully for each variant, a survey questionnaire is provided to gather user feedback for that variant.

\subsection{Survey questionnaire}
\label{subsec:surveyQ}
As we developed six CAPTCHA variants, it is essential to ask the right questions in order to determine which CAPTCHA achieves a better balance between security and usability. To evaluate usability, we designed a structured questionnaire. The survey questionnaire was developed based on usability dimensions commonly discussed in prior CAPTCHA studies \cite{chhabra2019design, kumar2022design, brodic2019exploring}, including readability, distortion, confidence, confusion, time consumption, annoyance, and comfort. However, the questionnaire is not claimed to be a standardized psychometric instrument. Future work may adopt or validate a standardized usability scale specifically for CAPTCHA evaluation. The questions are presented on Table \ref{tab:surveyQs}. The answer to every question from Q2 to Q9 has five options: \textbf{Never}, \textbf{Seldom}, \textbf{Sometimes}, \textbf{Often}, and \textbf{Always}, which are mapped to the numerical values ranging from 1 to 5 respectively. The questionnaire was used as a subjective usability instrument to complement objective measures such as success rate and response time. Each participant answered the questionnaire after interacting with multiple CAPTCHA samples from a given variant, not after viewing only a single CAPTCHA image. Therefore, frequency-based responses such as \textbf{Never}, \textbf{Seldom}, \textbf{Sometimes}, \textbf{Often}, and \textbf{Always} refer to the participant's experience across that variant.

\begin{table*}[h]
\centering
\resizebox{0.97\textwidth}{!}{%
\begin{tabular}{c|l|c}

\hline
\textbf{No.} & \multicolumn{1}{c|}{\textbf{Questions}}                & \textbf{Ideal Response} \\ \hline \hline
Q1 &
  \multicolumn{1}{l|}{\begin{tabular}[l]{@{}l@{}} How confident were you in your ability to solve previous CAPTCHA variant? (1 = Least \\ confident, 5 = Most confident
  \end{tabular}} &
  5 \\ \hline
Q2   & Was the CAPTCHA text distorted badly?         & Never (1)          \\ \hline
Q3   & Was the CAPTCHA readable?                     & Always (5)         \\ \hline
Q4   & Was the CAPTCHA prone to dictionary attack?   & Never (1)         \\ \hline
Q5   & Was the CAPTCHA variant solvable?             & Always (5)        \\ \hline
Q6   & Was the CAPTCHA variant confusing?            & Never (1)      \\ \hline
Q7   & Was the CAPTCHA time consuming to understand? & Seldom (2)         \\ \hline
Q8   & Was the CAPTCHA annoying?                     & Sometimes (3)      \\ \hline
Q9   & Was the CAPTCHA interesting to solve?         & Always (5)         \\ \hline
Q10 &
  \multicolumn{1}{l|}{\begin{tabular}[l]{@{}l@{}}How much comfortable you are of this CAPTCHA variant on the scale of 1 to 5? (1 = Least\\ comfortable, 5= Most comfortable)\end{tabular}} &
  5 \\ \hline
\end{tabular}%
}
\caption{Survey questions and their ideal responses for the standard variation}
\label{tab:surveyQs}
\end{table*}

\subsection{Participant recruitment}
\label{section20: participants recruitment}

We recruited 110 participants for the usability study. Recruitment was conducted both online, through social media announcements and university mailing lists, and in person on the campus of Shahjalal University of Science and Technology, Sylhet, Bangladesh~\cite{sust_profile} and through participants' personal networks. The study was reviewed and approved by the Institutional Review Board of Shahjalal University of Science and Technology~\cite{sust_profile}. All participants were native Bengali speakers, were at least 18 years of age, and provided informed consent through an in-application consent form before account creation. Participation was voluntary and no monetary compensation was offered. Participants were informed that the data collected interaction logs, and survey responses, would be anonymized and used solely for research purposes. No personally identifiable information was retained in the analysis dataset. None of the participants were students or direct collaborators of the research team. The sample comprised 86 male and 24 female participants.

\subsection{Usability Study}
\label{section21: operation}
During the study, when a user accesses the web application, a simple GUI is presented with all the instructions to carry out the required steps in both English and Bengali Language. Then, the user would need to complete the following steps:

\begin{enumerate}[label=(\roman*)]
    \item \textbf{Signup and login:} To access the web application, users would need to sign up by providing different information, including their email address, username, and password, as well as demographic details such as age, gender, internet experience, and education level on a web form. Next, the user would need to login by providing the login credentials such as username and password.

    \item \textbf{CAPTCHA challenge:} After logging in, users are presented with CAPTCHAs, as discussed previously. We have used six different CAPTCHA variants. Each participant instructed to solve three CAPTCHA samples successfully for each variant, in total solving 18 CAPTCHA to complete the study.

    \item \textbf{Survey questionnaire:} After successfully solving CAPTCHAs from a variant, a survey form is shown to the participant with 10 questions, as discussed in Section \ref{subsec:surveyQ}, regarding their experience with that variant. A CAPTCHA sample of the same variant will appear until a user solves three CAPTCHA samples of that variant. So, in total of 6 survey forms are filled by the participants to complete the study.
    
    \item \textbf{Choice list:} Finally, at the end of the study for each participants, they would be required to choose a CAPTCHA from the six variants to select the variant they were most comfortable with. This is an effort to collect important subjective preference data that complements the objective metrics we gathered earlier. By asking participants to select the CAPTCHA they felt most comfortable with, we can obtain a more comprehensive view of the overall user experience. This qualitative choice offers valuable insights into user acceptance and the potential for real-world adoption. From Table \ref{tab:variant_summary}, we can observe the distribution of participants' choices across the different CAPTCHA variants.
\end{enumerate}

\subsection{Data Analysis}
After all participants completed the study, we analyzed the collected data in different ways. We present out analysis In the following.
\subsubsection{Response time}
\label{subSection:Response Time}
The time it takes for a user to complete a CAPTCHA challenge successfully is defined as the response time. The less time it takes for a user to solve a CAPTCHA, the more usable the CAPTCHA is and vice versa. Our application recorded the response time for each CAPTCHA sample of each variation of the CAPTCHA. From Table \ref{table:successRateWithTime}, we can see the average response time for each variant.

\begin{table}[h] 
    \centering
    \begin{tabular*}{\linewidth}{@{\extracolsep{\fill}}c c c c c}
        \toprule
        \textbf{V} & \textbf{S} & \textbf{NS} & \textbf{SR} & \textbf{RT (s)} \\
        \midrule \midrule
        V1 & 279 & 217 & 56.25\% & 8.2581 \\ 

        V2 & 345 & 47 & 88.01\% & 7.7420 \\ 

        V3 & 372 & 40 & \textbf{90.29\%} & 9.9462 \\   

        V4 & 253 & 74 & 77.37\% & 9.1186 \\ 

        V5 & 171 & 28 & 85.93\% & \textbf{6.6901} \\   

        V6 & 273 & 36 & 88.35\% & 8.1832 \\ 
        \bottomrule
    \end{tabular*}
    \caption{Success rate and response time of different variants (V: Variation, S: Solved by human, NS: Not Solved by human, SR: Success Rate, RT: Response Time)}
    \label{table:successRateWithTime}
\end{table}

\subsubsection{Success rate}
The success rate refers to the percentage of the ratio between successfully solved CAPTCHAs and the total number of attempts \cite{kumar2022design}. The success rate is crucial as a low success rate can prevent legitimate users from accessing the website or application. Conversely, a higher success rate makes a CAPTCHA more usable. However, the success rate of CAPTCHA solving can vary depending on the type of CAPTCHA, the difficulty of the challenge, and the user’s knowledge and experience. Generally, more complex CAPTCHAs have lower success rates. 

Table \ref{table:successRateWithTime} shows that Variant 3 achieved the highest success rate of 90.29\%, while Variant 5 achieved the lowest average response time of 6.69 seconds. Most variants achieved success rates above 80\%, except Variant 1 and Variant 4

\subsection{Usability Performance: Normative Comparison}
Normative comparison involves evaluating something based on an ideal or standard that is considered the norm or the best possible variation. Normative comparisons are often used in quality control, product development, and performance evaluation to determine the degree to which an item meets a predetermined norm \cite{redmond2008normative}.

For the normative comparison, we have hypothesized an ideal CAPTCHA that is most usable by any standard. It is called the \textit{Standard CAPTCHA Variation}. We have idealized this CAPTCHA by setting the set of survey questions answers in the best possible way so that the usability of the CAPTCHA becomes maximum by any measure possible. The best possible answers to the questions are in Table \ref{tab:surveyQs}. The selection for the best answer has been made with extensive background study on CAPTCHA usability \cite{yan2008usability,Beheshti}.

\begin{figure}[h]
\centering
\includegraphics[width=0.6\linewidth]{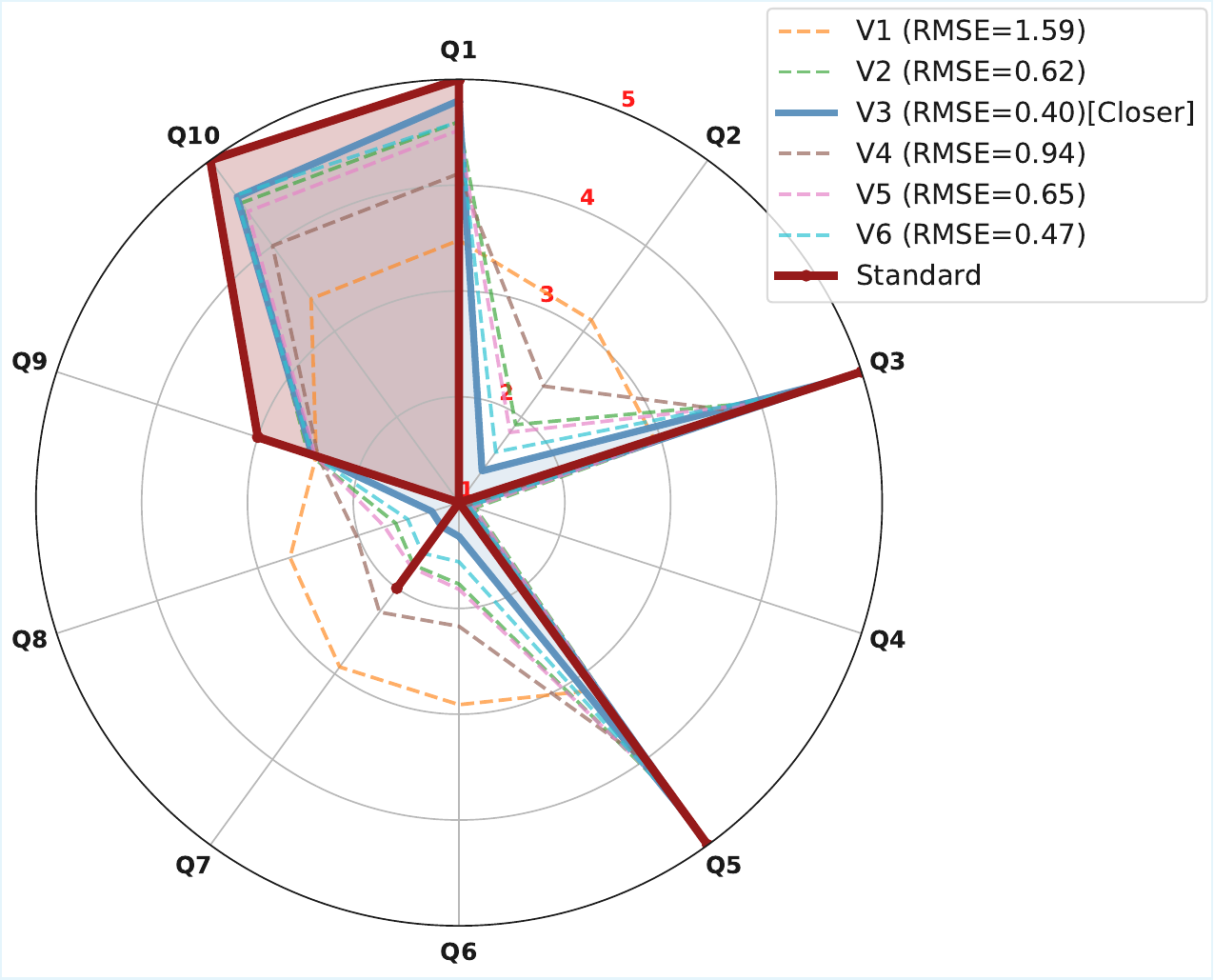}
\caption{Spider chart of standard responses and six variations across ten questionnaire items}
\label{fig:RadarChartIdealVairant}
\end{figure}

We can consider the CAPTCHA variant that is closest to the Standard CAPTCHA as the most usable. To calculate this `closeness', we collected numerical values from the answers (Table \ref{tab:surveyQs}) to each question in the Bengali CAPTCHA variations (V1-V6) and computed the mean score for each question. These mean values are reported in Table \ref{table:numericalMean} alongside the corresponding Standard CAPTCHA values.

\begin{table}[h]
    \centering
    \small
    
    \begin{tabular*}{\linewidth}{@{\extracolsep{\fill}} c c c c c c c c}
        \toprule
        \textbf{Questions} & \textbf{V1} & \textbf{V2} & \textbf{V3} & \textbf{V4} & \textbf{V5} & \textbf{V6} & \textbf{STC} \\ 
        \midrule \midrule
        
        Q1  & 3.47 & 4.53 & \textbf{4.80}  & 4.11 & 4.52 & 4.61 & 5 \\
        Q2  & 3.15 & 1.88 & \textbf{1.35} & 2.38 & 1.82 & 1.57 & 1 \\
        Q3  & 2.94 & 4.09 & \textbf{4.68} & 3.79 & 4.07 & 4.44 & 5 \\
        Q4  & 1.10 & 1.14 & \textbf{1.06} & \textbf{1.09} & 1.15 & 1.10 & 1 \\
        Q5  & 3.15 & 4.39 & \textbf{4.66} & 4.08 & 4.32 & 4.53 & 5 \\
        Q6  & 2.93 & 1.78 & \textbf{1.31} & 2.15 & 1.82 & 1.55 & 1 \\
        Q7  & 2.95 & 1.75 & 1.28 & 2.27 & \textbf{1.76} & 1.58 & 2 \\
        Q8  & \textbf{2.70} & 1.64 & 1.29 & 2.01 & 1.74 & 1.51 & 3 \\
        Q9  & 2.42 & \textbf{2.51} & 2.45 & 2.43 & 2.46 & 2.47 & 5 \\
        Q10 & 3.37 & 4.47 & 4.58 & 3.99 & 4.36 & \textbf{4.59} & 5 \\ 
        
        \bottomrule
    \end{tabular*}
    \caption{Numerical mean of survey response answers and Standard CAPTCHA (STC) answers}
    \label{table:numericalMean}
\end{table}

\subsubsection{Root Mean Squared Error (RMSE)}
For this analysis, we treated the 10 mean values (Q1-Q10) of each variant as a 10-dimensional vector and compared it to the Standard response vector using Root Mean Squared Error (\textbf{RMSE}). The purpose of this RMSE comparison is to assess how closely our CAPTCHA variants resemble the Standard CAPTCHA. Lower RMSE with a variant means it is closer to the standard CAPTCHA. We plotted the radar chart (Figure \ref{fig:RadarChartIdealVairant} on the standard response
profile of (Standard CAPTCHA) with six variations (V1-V6) across ten questionnaire items (Q1–Q10). The Standard condition is highlighted in red, while the closest matching variation (V3) computed by the lowest RMSE with the standard is highlighted in blue.

\begin{table*}[h]
\centering
\setlength{\tabcolsep}{4pt}
\renewcommand{\arraystretch}{1.15}

\begin{tabular*}{\linewidth}{@{\extracolsep{\fill}} l|cccccc}
\hline
\textbf{Metric} & \textbf{V1} & \textbf{V2} & \textbf{V3} & \textbf{V4} & \textbf{V5} & \textbf{V6} \\ \hline \hline
Distribution of Participant Choices & 4 & 32 & 8 & 19 & 24 & 23 \\ \hline
Correlation Coefficient with Ideal Response ($\rho$) & 0.5127 & 0.9746 & \textbf{0.9839} & 0.9362 & 0.9734 & 0.9837 \\ \hline
Success rate & 56.25\% & 88.01\% & \textbf{90.29}\% & 77.37\% & 85.93\% & 88.35\% \\ \hline
Response Time (in seconds) & 8.2581 & 7.7420 & 9.9462 & 9.1186 & \textbf{6.6901} & 8.1832 \\ \hline
RMSE (With Standard Questions) & 1.59 & 0.62 & \textbf{0.40} & 0.94 & 0.65 & 0.47 \\ \hline

\end{tabular*}

\caption{Summary of participant choices, correlation with the ideal response for six CAPTCHA variants.}
\label{tab:variant_summary}
\end{table*}

\subsubsection{Correlation Coefficient}
We can measure the similarity between each variant (V1-V6) using the correlation coefficient \cite{asuero2006correlation}. The correlation coefficient of each of the Variants with the hypothetical Standard CAPTCHA is presented in Table \ref{tab:variant_summary}. The correlation coefficient results indicate that Variant 3 ($\rho$ = 0.9839) achieves the highest correlation, closely followed by Variant 6 ($\rho$ = 0.9837), suggesting that these two variants align most strongly with the ideal response. Variants 5 ($\rho$ = 0.9734) and 2 ($\rho$ = 0.9746) also demonstrate high levels of consistency with the standard, while Variant 4 ($\rho$ = 0.9362) performs moderately. In contrast, Variant 1 ($\rho$ = 0.5127) shows the weakest correlation, indicating a considerable deviation from the expected response and consequently poorer usability. If we compare the two methods (RMSE and correlation coefficient) used to evaluate closeness to the Standard CAPTCHA and identify the most usable variant in Table \ref{tab:variant_summary}, both approaches yield consistent results, indicating that Variation 3 (V3) is the most usable CAPTCHA among the 6 variants.

\subsubsection{User Preference for Reduced Interaction Time}
Although statistical analysis indicated that variant V3 was the most usable, the majority of participants (32) expressed a preference for V2 (see Table \ref{tab:variant_summary}). Data from Table \ref{tab:variant_summary} shows that V2 had a lower mean response time of 7.74 seconds compared to V3 having a mean response time of 9.94 seconds. However, V3 achieved a higher success rate of 90.29\%, surpassing V2's success rate of 88.01\%. This highlights an interesting gap between performance and user preference. It appears that users prioritize minimizing the time to solve the challenge and resume their primary workflow over slightly better accuracy, as they preferred the version that saved them time, even if it meant experiencing a few more failures.

\subsubsection{Comparative analysis with established usability benchmarks}
To the best of our knowledge, this work is the first to design and implement a text CAPTCHA for Bengali language and to evaluate its usability through a user study. Consequently, no directly comparable benchmark or baseline exists for Bengali CAPTCHAs. Therefore, an exact comparison is not possible. Nevertheless, we compare our results with prior state-of-the-art usability studies of English text-based CAPTCHAs, given the similarities in task structure and interaction. As shown in Table \ref{table:baseline_time_success}, our Bengali CAPTCHA outperforms the reported baselines for text-based CAPTCHAs, achieving a lower response-time range and a higher success rate.
\begin{table}[h]
    \centering
    \begin{tabular*}{\linewidth}{@{\extracolsep{\fill}}l c c}
        \toprule
        \textbf{Baseline} & \textbf{Response Time (s)} & \textbf{Success Rate} \\
        \midrule \midrule
        Chellapilla et al.\cite{chellapilla2005designing} & 10 - 15 & 87.5\% \\
        Bursztein et al. \cite{bursztein2010good} & 9.8- 28.4 & 84\% \\
        Baecher et al. \cite{baecher2010captchas} & Not reported & 70-87\% \\
        Brodic et al. \cite{brodic2019captcha} & 13.38 - 22.95 & 0-77\%\footnotemark\\
        Searles et al. \cite{searles2023empirical} & 9 - 15.3 & 50-84\% \\
        \textbf{Bengali CAPTCHA} & \textbf{6.69 - 9.9} & \textbf{56.25 -90.29\%} \\
        \bottomrule
    \end{tabular*}
    \caption{Response time and success rate of human participants reported by other usability studies}
    \label{table:baseline_time_success}
\end{table}
\footnotetext{User under 35 years age 77\% probability and User over 35 were not able to solve any Text CAPTCHAs }
\section{Conclusion}
\label{sec: Conclusion}
In this work, we presented the first secure and usable Bengali Text CAPTCHA Scheme. It has six variants, each with distinct security properties. We implemented this design as a publicly accessible web application. We evaluated its security using an automated attack framework \cite{shibbir2024evaluating}, which combined image de-noising and OCR. The attack achieved a maximum average character recognition rate of only 20\% (Variant 6), confirming the substantial robustness of the scheme.

We conducted a usability study with two complementary analyses. The conventional success-rate metric yielded a 90.29\% overall success rate. Furthermore, we introduced a Normative Comparison method, which evaluated each variant by aligning its response distribution with an ideal usability profile. This analysis ranked Variant 3 as the most usable, a result consistent with its high success rate. We therefore conclude that Variant 3 represents the most secure and user-friendly text-based Bengali CAPTCHA, which can be successfully deployed for public use.

\vspace{2mm}
\noindent \textbf{Limitations and Future Work:} The current system has some limitations due to the complexities of rendering Bengali scripts. First, vowel and consonant conjuncts could not be incorporated, as the Python image library used would incorrectly segment them. Second, to keep randomized characters distinct and prevent automatic conjunct formation, participants had to enter a space after each character, which may have affected usability. Furthermore, the security of the proposed scheme should be interpreted in the context of current OCR-based attack capabilities. As Bengali OCR models improve, the robustness of the proposed CAPTCHA may change. However, the scheme does not depend exclusively on OCR limitations; it also incorporates multiple anti-preprocessing, anti-segmentation, and anti-recognition features.

\bibliographystyle{ACM-Reference-Format}
\bibliography{cas-refs}

\appendix
\section{Attacking Bengali CAPTCHAs}
\label{sec:attackBangladeshiCAPTCHA}

To validate the security claims of our developed Text-Based Bengali CAPTCHA, it was essential to go beyond theoretical evaluation and perform a practical, adversarial analysis. While the evaluation framework \cite{shibbir2024evaluating} estimates the strength of a CAPTCHA based on its security and vulnerability features, real world attackers do not rely on feature inspection they attempt to bypass the mechanism using automated tools systematically. Therefore, we deliberately attacked the system in a manner that closely resembles how modern bots operate in practice. This step was necessary to confirm whether the CAPTCHA, despite appearing secure under the framework, could still be compromised through realistic automated attacks. Such an attack-based validation ensures that the framework’s predictions are not only conceptual but also aligned with practical attack feasibility.

To validate the security of our Text-Based Bengali CAPTCHA, we attacked with a three-stage  \cite{shibbir2024evaluating} :
    \begin{enumerate}[label=(\roman*)]
        \item Pre-Processing
        \item OCR
        \item Post-Processing
    \end{enumerate}

\subsubsection{Pre-processing}
Its primary goal is to exaggerate information about characters in a particular picture and to reduce or erase interfering information \cite{chen2017survey}. The pre-processing stage is the most crucial stage because if a Text-CAPTCHA is properly pre-processed, then the segmentation and recognition can be done easily with modern technologies \cite{bansal2008breaking}. We have tried all the best possible combinations for our developed CAPTCHA to cancel out all the noises from the CAPTCHAs. All the best-chosen methods are inadequate for removing the security features from our Bengali CAPTCHA as presented in Table \ref{tab:preprocesedBengali}.

\begin{table*}[h]
\centering
\resizebox{0.9\linewidth}{!}{
    \begin{tabular*}{\linewidth}{@{\extracolsep{\fill}}c c c c c c c}
        \toprule
        \textbf{CAPTCHA} & \textbf{Binarization} & \textbf{Blurring} &
        \textbf{Erosion} & \textbf{Dilation} & \textbf{Gray-scale} & \textbf{P.P Output} \\
        \midrule
        \parbox[c]{4em}{\includegraphics[width=15mm,height=9mm]{img/captcha_variants/captcha1.png}} & 
        T.B.(128) & A.B.(3,1) & (2,2) & (3,2) & B2G &
        \parbox[c]{4em}{\includegraphics[width=15mm,height=9mm]{img/captcha_variants/NR_captcha1.png}} \\
        \parbox[c]{4em}{\includegraphics[width=15mm,height=9mm]{img/captcha_variants/captcha2.png}} & 
        OTSU(128) & A.B.(1,1) & (2,1) & (3,2) & B2G &
        \parbox[c]{4em}{\includegraphics[width=15mm,height=9mm]{img/captcha_variants/NR_captcha2.png}} \\
        \parbox[c]{4em}{\includegraphics[width=15mm,height=9mm]{img/captcha_variants/captcha3.png}} & 
        T.B.(100) & G.B.(13,13) & -- & (2,2) & B2G &
        \parbox[c]{4em}{\includegraphics[width=15mm,height=9mm]{img/captcha_variants/NR_captcha3.png}} \\
        \parbox[c]{4em}{\includegraphics[width=15mm,height=9mm]{img/captcha_variants/captcha4.png}} & 
        T.B.(100) & G.B.(13,13) & -- & (2,2) & B2G &
        \parbox[c]{4em}{\includegraphics[width=15mm,height=9mm]{img/captcha_variants/NR_captcha4.png}} \\
        \parbox[c]{4em}{\includegraphics[width=15mm,height=9mm]{img/captcha_variants/captcha5.png}} & 
        T.B.INV(105) & M.B.(5) & (1,1) & (2,2) & B2G &
        \parbox[c]{4em}{\includegraphics[width=15mm,height=9mm]{img/captcha_variants/NR_captcha5.png}} \\
        \parbox[c]{4em}{\includegraphics[width=15mm,height=9mm]{img/captcha_variants/captcha6.png}} & 
        T.B.(205) & G.B.(3,3) & (1,1) & (3,3) & B2G &
        \parbox[c]{4em}{\includegraphics[width=15mm,height=9mm]{img/captcha_variants/NR_captcha6.png}} \\
        \bottomrule
    \end{tabular*}
}
\caption{Pre-processing of selected CAPTCHA samples.}
\label{tab:preprocesedBengali}
\end{table*}

Next, we explain the different notations and parameters used in Table \ref{tab:preprocesedBengali}. They are described in detail in this work \cite{shibbir2024evaluating}

\begin{itemize}
    \item \textbf{T.B.(a)} denotes the \textbf{THRESH\_BINARY} operation.
    \item \textbf{T.B.INV(a) } denotes the \\ 
    \textbf{THRESH\_BINARY\_INVERSE} operation.
    \item \textbf{OTSU} denotes Otsu's method
    \item \textbf{G.B.(a,b)} denotes the \textbf{Gaussian Blur} operation where the values of a and b are the height and weight of the kernel respectively.
    \item \textbf{A.B.(a)} denotes the \textbf{Average Blur} operation where \textit{a} value is the same as G.B.

    \item \textbf{(a,b)} means the kernel height is a and the weight is b.
    \item \textbf{B2G} denotes BGR2GRAY, converting an Image to an Gray-scale Image.
\end{itemize}

We utilize the specific parameter configurations defined in our implementation’s attack script.

\begin{figure}[h]
\centering
\includegraphics[width=1\linewidth]{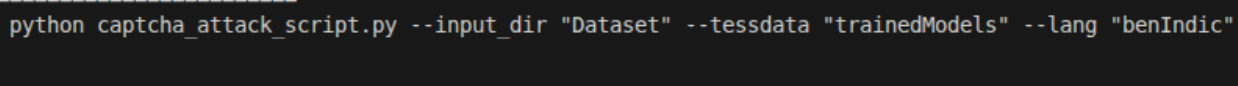}
\caption{Command to execute attack script}
\label{commandLinux}
\end{figure}

\begin{figure}[h]
\centering
\includegraphics[width=1\linewidth]{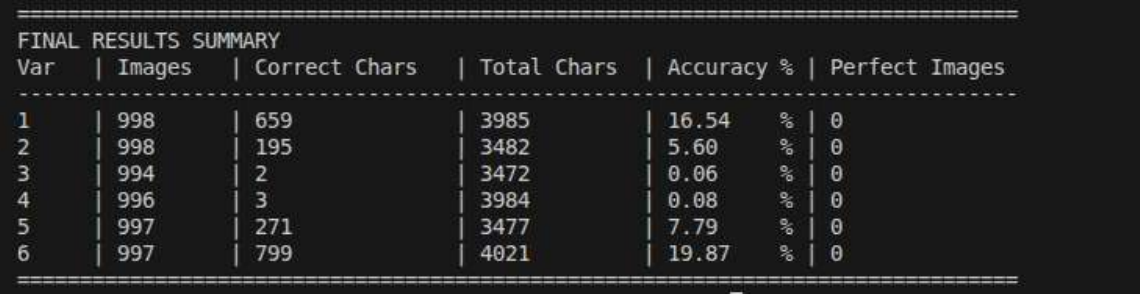}
\caption{Best Result from the pretrained model benIndic \cite{tesseractIndic}}
\label{implementationResultBenIndicBest}
\end{figure}

\begin{figure}[h]
\centering
\includegraphics[width=1\linewidth]{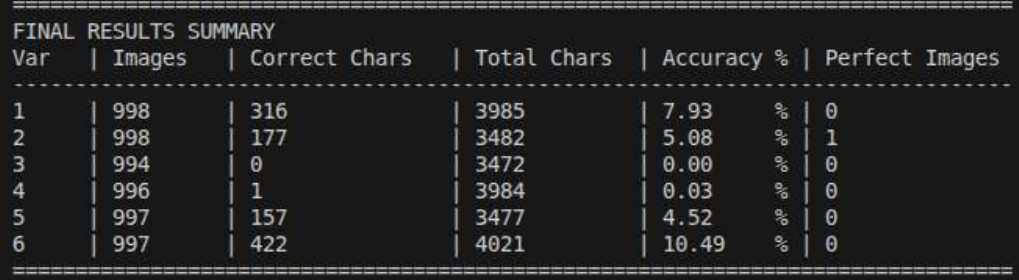}
\caption{Only one Bengali CAPTCHA sample of variation 2 out of 6000 samples was solved by two pretrained models\cite{vintasoft,tesseract-ocr-best}}
\label{vitanSoft}
\end{figure}
Figure \ref{vitanSoft} illustrates that only a single sample in Variation 2 was successfully solved by the pretrained models \cite{vintasoft, tesseract-ocr-best}, However they are showing lower average recognision rate than the pretrained model \cite{tesseractIndic}.

\subsubsection{OCR Selection}
Tesseract OCR is regarded as one of the most precise open-source OCR engines available \cite{openOcr}. The software looks at an image and finds individual characters or groups of characters, which are then turned into texts. It recognizes several text types, including handwriting and varied typefaces. A complete and detailed explanation of the Tesseract OCR engine is available at \cite{smith2007overview}.  

Tesseract was initially designed for English but has been expanded to recognize 100 other languages,  including Bengali \cite{tessdoc4}. The trained data file for each language is an archive file in a Tesseract-specific format. It contains several uncompressed component files needed by the Tesseract OCR process.  It is also called the Tesseract model. The official Tessaract repository provides three different kinds of trained data for each language.

Now, for the Bengali Language,  we must configure it with the Bengali pretrained data to recognize Bengali characters with this OCR for the Bengali language. The engine's algorithms are particular to the English alphabet, making supporting other structurally similar scripts easy by simply training the character set with the new scripts. Bengali also follows the same pattern of character segmentation. There are many versions of Bengali language pretrained model available, ranging its performance and accuracy \cite{tesseract-ocr-ben,tesseract-ocr-tableBenTrain,tesseractIndic, PyBnDataset1,githubTesseractBen, vintasoft,tesseract-ocr-ben-best, Tesseract-Ocr_fast}.

\subsection{Post-processing}
Post-processing is a crucial stage in attacking CAPTCHAs as it refines the recognition results and improves the solution’s accuracy. As discussed in \cite{shibbir2024evaluating}, the central aspect of post-processing is correcting recognition mistakes via error correction algorithms like pattern matching. Post-processing may accomplish this by comparing the recognized characters to a dictionary of known words, looking to the ground truth to find patterns, or using statistical models to anticipate the most probable nature based on the surrounding context. 
For our developed Bengali CAPTCHA, no post-processing is possible because the results after the recognition do not reveal any patterns, as the noises can not be removed with any standard noise removal techniques in the first place. Hence, recognition is not possible, resulting in no patterns.

\subsection{Attack Execution}
Our attack implementation is available at (./Attack/captcha\_attack\_script.py) within our open science repository\footnote{\url{https://github.com/neyamul-sbr/ArtifactsBengaliCAPTCHA}}. From Figure \ref{commandLinux}, we can see how we execute the attack script.
There, we have three primary environmental parameters:
\begin{itemize}
    \item \textbf{Input Directory (\texttt{--input\_dir}):} Specifies the path to the structured dataset containing six variations of the target captchas.
    \item \textbf{Model Path (\texttt{--tessdata}):} Overrides system defaults by pointing to a localized \texttt{trainedModels} directory. This ensures the use of specific model versions and maintains environment consistency across different systems.
    \item \textbf{Language Specification (\texttt{--lang}):} Designates the particular pretrained model, specifically targeting the Bengali language recognition file required for the attack execution. From Figure \ref{commandLinux}, we can see the  benIndic\cite{tesseractIndic} pretrained model was used for this command.
\end{itemize}
We evaluated the attack across nine distinct pre-trained models. Our results, illustrated in Figure \ref{implementationResultBenIndicBest}, indicate that while the \texttt{benIndic} model \cite{tesseractIndic} outperformed all alternative configurations for Bengali CAPTCHA recognition, it failed to achieve a complete successful recognition of any single CAPTCHA sample. This highlights the inherent difficulty of the target samples even when utilizing optimized OCR models.

\subsection{Discussion}
We have generated 1000 samples for our each CAPTCHA variants (6000 in total) and performed attack on them by the Pre-processing configuration of Table \ref{tab:preprocesedBengali}. The best results we found for trying to break it with 
attacks are shown in Table \ref{table:attackBengali}, where we can see that the average recognition rate from our developed CAPTCHA mechanism is highest 19.87\% for variation 2. 

The Average Prediction Rate is the percentage of an average number of characters recognized by performing the attack.

 \end{document}